\documentclass[fleqn,usenatbib]{mnras}
\usepackage{fix-cm}

\usepackage{newtxtext,newtxmath}

\usepackage[T1]{fontenc}

\DeclareRobustCommand{\VAN}[3]{#2}
\let\VANthebibliography\thebibliography
\def\thebibliography{\DeclareRobustCommand{\VAN}[3]{##3}\VANthebibliography}

\usepackage{graphicx}	
\usepackage{amsmath}	
\usepackage{verbatim}

\title[Statistical Insights into Flux and Photon Index Distributions of VHE FSRQs]{Statistical Insights into Flux and Photon Index Distributions of VHE FSRQs from \emph{Fermi}-LAT Observations}

\author[Malik et al.]{
Zahoor Malik,$^{1}$\thanks{E-mail:malikzahoor313@gmail.com}
Sikandar Akbar,$^{2}$\thanks{E-mail:darprince46@gmail.com}
Zahir Shah,$^{3}$\thanks{E-mail:shahzahir4@gmail.com}
Ranjeev Misra,$^{4}$
Athar A. Dar,$^{2,3}$\thanks{E-mail:ather.dar6@gmail.com}
Aaqib Manzoor,$^{5}$
\newauthor Sajad Ahanger,$^{2}$
Zeeshan Nazir,$^{3}$
Naseer Iqbal,$^{2}$
Seemin Rubab,$^{1}$
and Javaid Tantry$^{2}$
\\
$^{1}$Department of Physics, National Institute of Technology, Srinagar 190006, India.\\
$^{2}$Department of Physics, University of Kashmir, Srinagar 190006, India.\\
$^{3}$Department of Physics, Central University of Kashmir, Ganderbal 191201, India.\\
$^{4}$Inter-University Centre for Astronomy and Astrophysics, Post Bag 4, Ganeshkhind Pune 411007, India.\\
$^{5}$Indian Institute of Astrophysics, Bangalore 560034, India.
}

\date{Accepted XXX. Received YYY; in original form ZZZ}

\pubyear{\the\year{}}

\begin{document}
\label{firstpage}
\pagerange{\pageref{firstpage}--\pageref{lastpage}}
\maketitle

\begin{abstract}
This study examines the flux and photon index distributions of 11 Very High Energy (VHE) Flat Spectrum Radio Quasars (FSRQs) using over 16 years of \emph{Fermi}-LAT $\gamma$-ray data. The distributions reveal double lognormal profiles in both flux and index, primarily in the 3-day and 7-day binnings, supporting the ``two-flux-state hypothesis" for blazars. These profiles, which become insignificant at 30-day binning, suggest that shorter timescales are better at capturing distinct states, while longer timescales smooth out shorter variations. Most VHE FSRQs exhibit a ``harder-when-brighter" trend, where the photon index decreases during high-flux states, suggesting efficient particle acceleration and possibly reduced radiative cooling. In contrast, two sources display a ``softer-when-brighter" behavior, likely due to enhanced radiative cooling in high photon density environments. Additionally, we observe that the Spearman rank correlation between flux and photon index strengthens with increasing time bin sizes, indicating more pronounced correlations over longer timescales. This possibly indicates that, on shorter timescales, flux variations are driven by a combination of photon index changes and normalization effects. Averaging flux over longer durations minimizes the effect of normalization variation, thereby enhancing the observed correlation. We also compare the flux and index distributions of VHE and non-VHE FSRQs, emphasizing the differences in their variability and emission patterns.
\end{abstract}




\section{Introduction}
\label{sec:intro}

Blazars are an extreme subclass of active galactic nuclei (AGN), distinguished by their intense, rapidly varying emissions across the entire electromagnetic spectrum, from radio to $\gamma$-rays. The intense emission observed in blazars is primarily attributed to relativistic jets directed nearly along the line of sight of observer \citep{1995PASP..107..803U}. These jets, moving at velocities close to the speed of light, induce Doppler boosting, in which the radiation from the jet is significantly amplified due to the relativistic motion of the emitting particles
 \citep{1979ApJ...232...34B, 1995PASP..107..803U}. This amplification leads to an increase in both the observed flux and variability, making blazars highly luminous across a broad range of wavelengths, from radio to $\gamma$-rays \citep{1998MNRAS.299..433F}. Blazars are classified into two primary types based on their optical emission-line properties: flat-spectrum radio quasars (FSRQs), which exhibit strong and broad emission lines, and BL Lacertae (BL Lac) objects, which have weak or even absent emission lines \citep{1995PASP..107..803U, 2003ApJ...585L..23F}. This distinction is believed to reflect differences in their jet environments, accretion mechanisms, and possibly the external photon fields that contribute to the observed radiation of blazars.

A subset of FSRQs has been identified as very high-energy (VHE) $\gamma$-ray emitters, with photon energies exceeding 100 GeV. These detections have been made using ground-based Cherenkov telescopes, including \emph{MAGIC}, \emph{VERITAS}, and \emph{H.E.S.S.} \citep{2006A&A...457..899A, 2006APh....25..391H, 2012APh....35..435A}. To date, only 11 FSRQs have been confirmed as VHE sources, highlighting their rarity within the broader population of blazars \footnote{\url{http://tevcat.uchicago.edu/}}. It is important to note that VHE telescopes are not survey instruments and can only observe sources that are visible under favorable conditions and fit within their scheduled priorities. These constraints significantly impact the detection rate of VHE FSRQs, emphasizing the need for targeted follow-up observations. The detected VHE emission in these sources is likely the result of complex processes, including inverse Compton scattering, where high-energy electrons interact with external photon fields, as well as hadronic interactions, such as proton-photon interactions, which can lead to the production of secondary $\gamma$-rays through processes like pion decay \citep{2013ApJ...768...54B}.

FSRQs often display characteristic trends in their flux-index relationship, such as the ``harder-when-brighter" or ``softer-when-brighter" behaviors \citep{2019MNRAS.484.3168S, 2022MNRAS.514.4259M, AKBAR2025438}. These trends reflect the interplay between particle acceleration and cooling processes within the relativistic jet and provide valuable information into the physical conditions driving $\gamma$-ray emission. Exploring these variability patterns, particularly in VHE-emitting FSRQs, can shed light on the mechanisms underlying high-energy radiation and the distinguishing features of VHE-emitting blazars compared to the broader blazar population.

Blazar variability, particularly in the $\gamma$-ray band, has been shown to follow a lognormal distribution, a pattern indicative of underlying multiplicative processes such as turbulence and particle cascades within relativistic jets, as discussed in \citet{2018RAA....18..141S} and references therein. This lognormal behavior suggests that the flux variability in blazars is not random but arises from dynamic processes within the jet environment. Energy dissipation through turbulence, driven by chaotic motions in the jet, and particle cascades, where high-energy particles interact with external photon fields leading to $\gamma$-ray emission, are key mechanisms behind this variability \citep{2010MNRAS.402..497G}. Furthermore, variations in the photon index, which describes the spectral slope, are often correlated with changes in flux, reinforcing the connection between particle acceleration mechanisms and the energy distribution in the jet \citep{2010ApJ...716...30A}. Several studies have demonstrated that the flux distributions of blazars, across both high and low-redshift sources, often follow lognormal patterns. However, other statistical fits, such as double lognormal distributions, have also been explored, as they can sometimes provide better descriptions of the flux variability, depending on the specific blazar population under study \citep{2016ApJ...822L..13K, 2019MNRAS.484.3168S, 2020MNRAS.491.1934K}. These statistical behaviors are critical to understanding the physical processes driving blazar variability and may help distinguish different subclasses of blazars, such as VHE-emitting FSRQs, from non-VHE counterparts.

This study uses over 16 years of continuous $\gamma$-ray observations from the \emph{Fermi}-LAT to perform a detailed statistical analysis of the flux and photon index distributions in VHE FSRQs. We analyze the histograms of $\gamma$-ray flux and photon index, applying different profiles to model the data. By fitting these profiles, we aim to understand patterns that may reveal the underlying processes of energy dissipation in the jets of these objects, such as particle acceleration, cooling, and the effects of magnetic fields. The study also investigates the correlation between flux and photon index across different timescales, enabling us to distinguish between short-term and long-term variability. Furthermore, by comparing the variability and spectral properties of VHE FSRQs with those of non-VHE FSRQs, we identify likely differences, which may indicate distinct physical conditions in the jets of VHE FSRQs that facilitate the production of VHE $\gamma$-rays. While the focus of this study is on statistical analysis, the results may provide valuable clues about the physical mechanisms that drive the variability and high-energy emission in these sources, advancing our understanding of the conditions necessary for VHE $\gamma$-ray emission in blazars.

This paper is structured as follows: Section \ref{sec:overview} gives an overview of VHE FSRQs, discussing their VHE detections and corresponding $\gamma$-ray emission. Section \ref{sec:data} explains the data analysis process, focusing on how we prepared and processed the \emph{Fermi}-LAT data for our study. Section \ref{sec:results} presents the results, examining the fitting profiles and the flux-index distribution correlations of VHE FSRQs, and comparing them to the broader FSRQ population. Finally, Section \ref{sec:summary} provides a summary of the key findings.

\section{Overview of VHE FSRQs} \label{sec:overview}

As of today, 11 FSRQs have been identified as VHE emitters. Most of these detections occurred during intense flaring episodes when their $\gamma$-ray flux increased significantly compared to average flux. These FSRQs have been identified primarily through \emph{Fermi}-LAT’s monitoring of their $\gamma$-ray flares, in conjunction with observations from ground-based VHE telescopes like \emph{MAGIC}, \emph{VERITAS}, and \emph{H.E.S.S.} etc. The sources listed below include details of their discovery dates, and key information linking \emph{Fermi}-LAT observations with subsequent VHE detections.

\textbf{S3 0218+35:} S3 0218+35, also known as 4FGL J0221.1+3556, was detected at VHE by the MAGIC telescope in July 2014 after experiencing a violent flare observed by the \emph{Fermi}-LAT \citep{2016A&A...595A..98A}. Located at R.A.: 02$^\mathrm{h}$21$^\mathrm{m}$05.5$^\mathrm{s}$, Dec.: +35$^\circ$56$'$14$''$ (J2000), this blazar is notable for being gravitationally lensed, with a redshift of $\rm z = 0.944 \pm 0.002$. The \emph{MAGIC} telescope observed the source for 3.5 hours from July 23 to July 26, 2014, estimating the VHE flux to be about 15$\%$ of the Crab Nebula's flux in the 100–200 GeV range. This made S3 0218+35 one of the most distant blazars detected at VHE by Cherenkov telescopes. \emph{Fermi}-LAT had already identified $\gamma$-ray flares from this source in 2012, observing a characteristic 11.46-day delay between the direct and lensed components of the flare, a result of gravitational lensing \citep{2014ApJ...782L..14C}. On July 13 and 14, 2014, \emph{Fermi}-LAT detected another flare with exceptionally hard spectra \citep{2014ATel.6316....1B}. The daily averaged fluxes were \(6.5 \pm 1.4 \times 10^{-7} \, \mathrm{photons \, cm^{-2} \, s^{-1}}\) and \(6.7 \pm 1.5 \times 10^{-7} \, \mathrm{photons \, cm^{-2} \, s^{-1}}\), with photon indices of \(1.4 \pm 0.1 \) and \(1.6 \pm 0.1 \) respectively. This prompted the \emph{MAGIC} observations, which led to the first VHE detection of a gravitationally lensed blazar.

\textbf{PKS 0346-27:} PKS 0346-27, also known as 4FGL J0348.5-2749, was observed in November 2021 by the \emph{H.E.S.S.} array, which detected a $>5\sigma$ excess in the VHE $\gamma$-ray band from its direction \citep{2021ATel15020....1W}. This FSRQ is located at R.A.: 03$^\mathrm{h}$48$^\mathrm{m}$38$^\mathrm{s}$, Dec.: -27$^\circ$49$'$14$''$ (J2000), with a redshift of $\rm z = 0.991$. The \emph{H.E.S.S.} observations on November 3, 2021, revealed a very soft photon spectral index greater than $4$. Prior to the VHE detection, the \emph{Fermi}-LAT had observed a significant increase in $\gamma$-ray activity from PKS 0346-27 on November 2, 2021. The daily averaged $\gamma$-ray flux (E$>$100 MeV) reached \(1.8 \pm 0.2 \times 10^{-6} \, \mathrm{photons \, cm^{-2} \, s^{-1}}\), marking a 200-fold increase compared to its average flux in the fourth \emph{Fermi}-LAT catalog (4FGL). The source also showed a photon index of \(1.8 \pm 0.1 \), which was significantly lower than its catalog value of \(2.5 \pm 0.1 \). Earlier $\gamma$-ray activity was noted on February 20, 2020 \citep{2020ATel13521....1M}, but the November 2021 flare marked the first significant detection of PKS 0346-27 at VHE. This demonstrates the importance of \emph{Fermi}-LAT’s role in identifying flaring episodes that lead to ground-based VHE detections.

\textbf{PKS 0736+017:} PKS 0736+017 (4FGL J0739.2+0137, R.A.: 07$^\mathrm{h}$39$^\mathrm{m}$18$^\mathrm{s}$, Dec.: +01$^\circ$37$'$05$''$ (J2000)) was detected at VHE by \emph{H.E.S.S.} on the night of February 19, 2015, during a low energy $\gamma$-ray flare that had been observed by \emph{Fermi}-LAT \citep{2020A&A...633A.162H}. \emph{Fermi}-LAT detected the flare on February 18, 2015, marking the onset of a significant increase in $\gamma$-ray activity. Strictly simultaneous observations with \emph{Fermi}-LAT during the \emph{H.E.S.S.} detection reported a flux of \(4.57 \pm 1.52 \times 10^{-6} \, \mathrm{photons \, cm^{-2} \, s^{-1}}\) and a photon index \(2.43 \pm 0.33 \). The \emph{Fermi}-LAT flare exhibited a doubling timescale of approximately six hours. With a redshift of $z = 0.189$, PKS 0736+017 is one of the nearest FSRQs detected at VHE, highlighting the importance of simultaneous observations from both \emph{Fermi}-LAT and \emph{H.E.S.S.} in capturing these short-lived, high-energy $\gamma$-ray events.

\textbf{PKS 0903-57:} PKS 0903-57 (4FGL J0904.9-5734, R.A.: 09$^\mathrm{h}$04$^\mathrm{m}$53.70$^\mathrm{s}$, Dec.: $-$57$^\circ$35$'$05.86$''$ (J2000)) was first identified as a radio source by \citet{1964AuJPh..17..340B}. The presence of a bright foreground star ($G \sim 16$) located merely 0$''$.67 away had long obscured its classification. However, recent observations suggest that PKS 0903-57 is a rare, low-luminosity VHE FSRQ, characterized by a conspicuous broad-line emission and a luminous continuum \citep{2024A&A...691L...5G}. PKS 0903-57 has recently attracted considerable interest due to strong flaring episodes observed since 2020 in high-energy and VHE $\gamma$-rays. On April 1, 2020, a flaring state at high-energy $\gamma$-rays was reported by \emph{AGILE} \citep{2020ATel13602....1L}, and both high-energy and VHE activity was detected by \emph{Fermi}-LAT \citep{2020ATel13604....1B}. Follow-up observations with the \emph{H.E.S.S.} array were planned; however, unfavorable weather conditions initially hampered prompt observations. Nevertheless, two observation runs totaling 49 minutes were successfully carried out on April 13, 2020, leading to a significant detection at a level exceeding 25$\sigma$ in a preliminary real-time analysis \citep{2020ATel13632....1W}. Further flaring episodes have been observed from the source in subsequent monitoring campaigns \citep{2022ATel15666....1L}.

\textbf{TON 0599:} Ton 0599 (also known as 4C +29.45, RGB J1156+292, 4FGL J1159.5+2914) was detected for the first time in VHE by the \emph{MAGIC} telescope on December 15, 2017 \citep{2017ATel11061....1M}. The object, located at R.A.: 11$^\mathrm{h}$59$^\mathrm{m}$31.83$^\mathrm{s}$, Dec.: +29$^\circ$14$'$43.83$''$ (J2000), was observed for around one hour, leading to a detection with a significance of about $10\sigma$. The estimated VHE flux was approximately \(0.15 \times 10^{-9} \, \mathrm{photons \, cm^{-2} \, s^{-1}}\), corresponding to about $0.3$ Crab Units above 100 GeV, with a soft photon spectrum. Ton 0599, an FSRQ at redshift $z = 0.72449$, was in an elevated state across multiple wavelengths (optical to $\gamma$-ray) since October 2017. Between October 28 and November 5, 2017, \emph{Fermi}-LAT observed bright flaring activity from the source, with daily $\gamma$-ray flux peaking at \(2.3 \pm 0.3 \times 10^{-6} \, \mathrm{photons \, cm^{-2} \, s^{-1}}\) on October 31, with a photon index of \(1.9 \pm 0.1 \) \citep{2017ATel10931....1C}. {During this time, \emph{VERITAS} also detected VHE emission from Ton 0599, confirming the activity across different instruments and energy bands \citep{2017ATel11075....1M, 2022ApJ...924...95A}.

\textbf{4C +21.35:} VHE $\gamma$-ray emission from the FSRQ 4C +21.35 (also known as PKS 1222+21, 4FGL J1224.9+2122; $\rm z = 0.432$) was detected by the \emph{MAGIC} during a brief observation of approximately 0.5 hours on June 17, 2010 \citep{2011ApJ...730L...8A}. The source is located at R.A.: 12$^\mathrm{h}$24$^\mathrm{m}$54.4$^\mathrm{s}$, Dec.: +21$^\circ$22$'$46$''$ (J2000). This detection coincided with significant high-energy MeV/GeV $\gamma$-ray activity observed by the \emph{Fermi}-LAT. In the MeV/GeV energy range, the source exhibited a notable flare that lasted about 3 days, peaking on June 18, 2010. During the period of the \emph{MAGIC} observation, the \emph{Fermi} integral flux was measured at \(6.5 \pm 1.9 \times 10^{-6} \, \mathrm{photons \, cm^{-2} \, s^{-1}}\), with a photon index of \(1.95 \pm 0.21 \). The \emph{VERITAS} collaboration also detected VHE $\gamma$-rays from PKS 1222+21 during observations between February 26 and March 10, 2014, following a bright GeV flare reported by \emph{Fermi}-LAT \citep{2014ATel.5981....1H, 2015arXiv150103554C}. A $\gamma$-ray excess was observed at $\sim 6\sigma$ significance, with an integral flux above 100 GeV of \(1.4 \pm 0.3 \times 10^{-11} \, \mathrm{photons \, cm^{-2} \, s^{-1}}\), $\sim 3\%$ of the Crab Nebula flux. The coordination between VHE telescopes  and \emph{Fermi}-LAT observations highlights the dynamic behavior of PKS 1222+21 during these active phase.

\textbf{3C 279:} The FSRQ 3C 279 (4FGL J1256.1-0547, $z = 0.5362$, R.A.: 12$^\mathrm{h}$56$^\mathrm{m}$11.1$^\mathrm{s}$, Dec.: +05$^\circ$47$'$22$''$ (J2000)) was detected at VHE $\gamma$-rays by the \emph{MAGIC} telescope during bright optical flares in 2006 \citet{2008Sci...320.1752M} and 2007 \citep{2011A&A...530A...4A}. However, no VHE $\gamma$-ray detection was reported for a long period despite multiple observations \citep{2014A&A...564A...9H, 2014A&A...567A..41A, 2016AJ....151..142A}. In the high-energy $\gamma$-ray regime, 3C 279 has been observed with both \emph{EGRET} \citep{1992ApJ...385L...1H} and \emph{Fermi}-LAT \citep{2009ApJ...700..597A}, with \emph{Fermi}-LAT detecting multiple flares in recent years, some of which were followed up with Cherenkov telescopes \citep{2014HEAD...1410611E, 2022ApJ...924...95A}. In April 2014 and June 2015, 3C 279 exhibited strong high energy $\gamma$-ray outbursts, with fluxes exceeding \(10^{-5} \, \mathrm{photons \, cm^{-2} \, s^{-1}}\) on hour-long timescales \citep{2015ApJ...807...79H, 2015ApJ...803...15P}. Follow-up observations with \emph{H.E.S.S.} did not detect VHE $\gamma$-rays in 2014, but a significant detection was made in 2015 \citep{2019A&A...627A.159H}. As part of the \emph{H.E.S.S.} Target of Opportunity program, 3C 279 was again observed multiple times in 2017 and 2018 following high optical and high energy $\gamma$-ray states. In January 2018, \emph{H.E.S.S.} detected an unexpected VHE flare at the end of an MeV-GeV flaring state \citep{2018ATel11239....1N}, while in June 2018, the decaying phase of a strong Fermi-LAT flare was continuously monitored for several nights, leading to a highly significant VHE detection \citep{2019ICRC...36..668E}. The source remains a key object of interest for studying $\gamma$-ray emission in FSRQs.

\textbf{OP 313:} The very distant FSRQ OP 313 (4FGL J1310.5+3221, $\rm z = 0.997$, R.A.: 13$^\mathrm{h}$10$^\mathrm{m}$28.6638$^\mathrm{s}$, Dec.: +32$^\circ$20$'$43.783$''$ (J2000)) has been under observation by the Large-Sized Telescope (\emph{LST-1}) since November 2023. Following an alert from \emph{Fermi}-LAT and optical facilities, renewed high-energy (E$>$100 MeV) activity was detected. Target of Opportunity observations from \emph{LST-1} between December 11 and 14, 2023, confirmed a significant VHE detection, with an integrated flux above 100 GeV at $15\%$ of the Crab Nebula's flux \citep{2023ATel16381....1C}. \emph{Fermi}-LAT observations showed the $\gamma$-ray flux peaking at \(1.8 \pm 0.2 \times 10^{-6} \, \mathrm{photons \, cm^{-2} \, s^{-1}}\) on November 24, 2023, with a hard photon index of \(1.80 \pm 0.06 \), much harder than the 4FGL catalog value \citep{2023ATel16356....1B}. This was the highest daily flux recorded for OP 313, with the previous flare reported in June 2022.

\textbf{B2 1420+32:} The \emph{MAGIC} collaboration reports the detection of VHE $\gamma$-ray emission from the blazar B2 1420+32 (OQ 334, 4FGL J1422.5+3223) with R.A.: 14$^\mathrm{h}$22$^\mathrm{m}$30.83$^\mathrm{s}$, Dec.: +32$^\circ$23$'$10.44$''$ (J2000), located at redshift $\rm z = 0.682$ \citep{2020ATel13412....1M, 2021A&A...647A.163M}. On January 21, 2020, during 1.6 hours of observation, a $13\sigma$ detection was achieved, with the VHE flux estimated to be about $15\%$ of the Crab Nebula’s flux above 100 GeV. The \emph{MAGIC} observation was triggered by \emph{Fermi}-LAT’s report of flaring activity \citet{2020ATel13382....1C} in GeV $\gamma$-rays. On December 30 and 31, 2019, the source exhibited significant activity, with daily averaged $\gamma$-ray fluxes of \(0.9 \pm 0.1 \times 10^{-6} \, \mathrm{photons \, cm^{-2} \, s^{-1}}\) and \(1.7 \pm 0.2 \times 10^{-6} \, \mathrm{photons \, cm^{-2} \, s^{-1}}\), respectively. This represents flux increase of approximately 110 and 210 times the eight-year average flux from the 4FGL catalog. The corresponding photon indices for these dates were \(1.9 \pm 0.1 \) and \(2.1 \pm 0.1 \).

\textbf{PKS 1441+25:} The \emph{MAGIC} collaboration reported the first discovery of VHE $\gamma$-ray emission from the FSRQ PKS 1441+25 (4FGL J1443.9+2501, R.A.: 14$^\mathrm{h}$43$^\mathrm{m}$56.9$^\mathrm{s}$, Dec.: +25$^\circ$01$'$44$''$ (J2000)), at redshift $\rm z = 0.939$ \citep{2015ATel.7416....1M, 2015ApJ...815L..23A}. Observations were conducted with the \emph{MAGIC} telescope during April 17–19, 2015, with detections reaching more than $6\sigma$ and $11\sigma$ on consecutive nights. The VHE flux above 80 GeV was approximately \(8 \times 10^{-11} \, \mathrm{photons \, cm^{-2} \, s^{-1}}\) ($16\%$ of Crab Nebula flux). Triggered by enhanced multi-wavelength activity, including optical, X-ray, and $\gamma$-ray frequencies \citep{2015ATel.7402....1P}, \emph{Fermi}-LAT detected a $\gamma$-ray flare from March 21 to April 15, 2015, with a flux of \(3.8 \pm 0.3 \times 10^{-7} \, \mathrm{photons \, cm^{-2} \, s^{-1}}\), and a photon index of \(1.93 \pm 0.07 \). This flux was significantly higher than the catalog value from the third Fermi-LAT catalog. Subsequently, \emph{VERITAS} detected VHE $\gamma$-ray emission from PKS 1441+25 during observations conducted on the night of April 21, 2015 \citep{2015ATel.7433....1M, 2015ApJ...815L..22A}.

\textbf{PKS 1510-089:} The source PKS 1510-089 (4FGL J1512.8-0906), with coordinates R.A.: 15$^\mathrm{h}$12$^\mathrm{m}$50.5$^\mathrm{s}$, Dec.: -09$^\circ$06$'$00$''$ (J2000), located at redshift $\rm z = 0.361$, was firstly observed with the \emph{H.E.S.S.} array during high states in the optical and GeV bands \citep{2013A&A...554A.107H}. VHE $\gamma$-rays were detected with a $9.2\sigma$ significance in 15.8 hours of data taken during March and April 2009. A VHE integral flux of \(1.0 \pm 0.2_{\text{stat}} \pm 0.2_{\text{sys}} \times 10^{-11} \, \mathrm{photons \, cm^{-2} \, s^{-1}}\) was measured. During the same period, the average integral flux in the 200 MeV–100 GeV range was \(1.26 \pm 0.03 \times 10^{-6} \, \mathrm{photons \, cm^{-2} \, s^{-1}}\), indicating simultaneous GeV and VHE activity. Subsequent observations were guided by multiwavelength flares, leading to a detection with \emph{MAGIC} during another high state of high energy $\gamma$-ray emission in 2012 \citep{2014A&A...569A..46A}. Systematic monitoring efforts at VHE $\gamma$-rays commenced only later \citep{2019Galax...7...41Z}, ultimately resulting in the detection of PKS 1510$-$089 in VHE $\gamma$-rays with \emph{MAGIC} during low high energy $\gamma$-ray states \citep{2018A&A...619A.159M}. In 2016, continuous monitoring with \emph{H.E.S.S.} led to the detection of a strong VHE $\gamma$-ray flare that was not preceded by significant multiwavelength activity. This event was subsequently followed up with \emph{MAGIC}, enabling, for the first time, investigations of sub-hour variability timescales in VHE $\gamma$-rays for this source \citep{2021A&A...648A..23H}.

\section{\emph{Fermi} LAT Data and Analysis} \label{sec:data}

The \emph{Fermi} Large Area Telescope (LAT), operating since 2008, is a space-based $\gamma$-ray observatory that monitors the sky from 20 MeV to more than 300 GeV \citep{2009ApJ...697.1071A}. With its broad field of view ($\sim2.4$ steradians) and continuous sky coverage, \emph{LAT} observes the entire sky every three hours, making it a powerful tool for time-domain studies of transient and variable $\gamma$-ray sources, such as blazars. This long-term monitoring, coupled with its high duty cycle, makes \emph{LAT} essential for studying variability and identifying flaring episodes in these sources.

For the analysis, we utilized data from the \emph{Fermi}-LAT Light Curve Repository \footnote{\url{https://fermi.gsfc.nasa.gov/ssc/data/access/lat/LightCurveRepository/}}, which contains flux-calibrated light curves for over 1,500 variable sources (variability index $> 21.67$) from the 4FGL-DR2 catalog \citep{2023ApJS..265...31A}. The light curves cover the entire mission, with data binned into 3-day, 7-day, and 30-day intervals, providing a high cadence and long-term monitoring of the $\gamma$-ray sky. These properties make the LCR an essential tool for time-domain astronomy. The final results include the best-fit flux or upper limit, along with the spectral shape for both fixed and variable photon index models, ensuring a comprehensive characterization of source variability and spectral evolution.

\section{Results} \label{sec:results}

\subsection{Gamma-Ray Light Curves and Statistical Distribution Analysis of VHE FSRQs}

In this study, we utilize $\gamma$-ray light curves of FSRQs spanning the energy range of 0.1–100 GeV, obtained from the \emph{Fermi}-LAT LCR. These light curves are presented in three different time binnings: 3-day, 7-day, and 30-day intervals. Alongside the flux values, we also incorporate the corresponding spectral index values to provide a comprehensive view of the spectral behavior of these sources across the \emph{Fermi}-LAT energy band. Figure \ref{fig:lc} displays the 7-day binned $\gamma$-ray light curves for the 11 VHE FSRQs, along with their corresponding spectral indices. These figures reveal significant variability in both the flux and spectral index over time, indicating a dynamic and complex emission mechanism. The red vertical lines in the figures highlight the times when VHE emission was detected, allowing us to see the correlations between heightened $\gamma$-ray flux and VHE activity. These observations suggest that the VHE detections often coincide with periods of increased $\gamma$-ray flux, providing additional insights into the potential physical processes responsible for these high-energy events.

\begin{figure*}
    \centering
    \includegraphics[scale=0.33,angle=0]{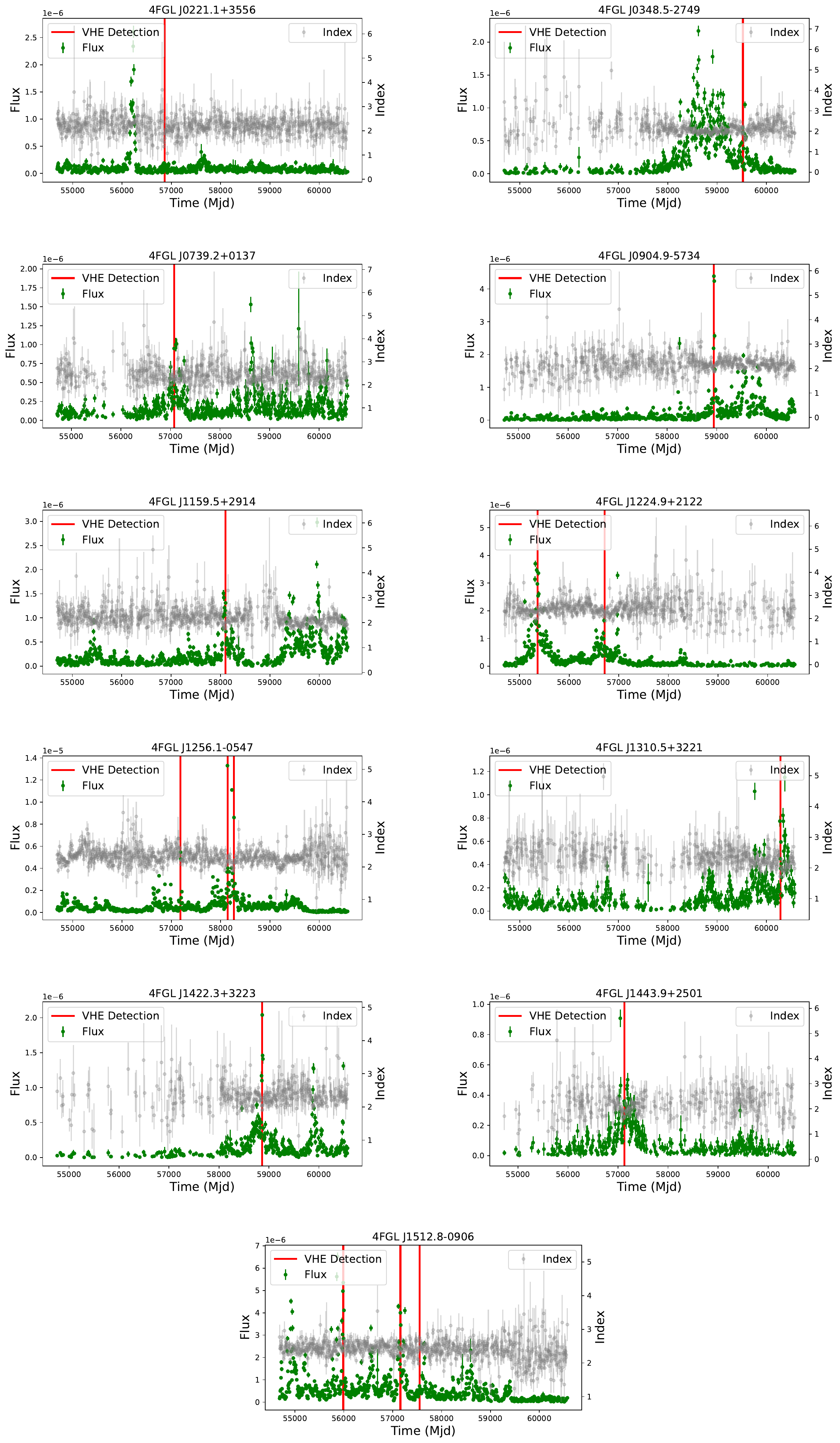}
    \caption{$\gamma$-ray light curves of 11 VHE FSRQs. Green points represent the photon flux in the 0.1--100 GeV energy range (measured in units of photons cm$^{-2}$ s$^{-1}$), while grey points correspond to the photon index. Red vertical lines mark the times at which the source was detected in VHE.\\
    \textit{Note:} At times, closely spaced detections by multiple instruments cause overlapping red lines in the figure. However, all detections are included in the text.}
    \label{fig:lc}
\end{figure*}


\subsubsection{Data Refinement and Normality Tests}

To ensure the robustness of our analysis, several refinement criteria were applied to the $\gamma$-ray photon flux and photon index data. We first filtered out data points with TS $<$ 4, which corresponds to a $2\sigma$ detection threshold. This ensures that only statistically significant measurements are considered in the analysis. Additionally, we applied strict criteria on the flux and index values, excluding those with large uncertainties. Specifically, we only retained flux and index values where the ratio of the value to its error exceeded 2.

To explore the statistical distributions of the flux and spectral index values, we performed normality tests using the Anderson-Darling (AD) test. This test was conducted both on the linear and log scales of flux and index data to determine whether it follows a normal or log-normal distribution. If the AD test indicated that the data followed a normal or log-normal distribution, we proceeded with the corresponding fit. In cases where neither normal nor log-normal distributions provided a good fit, we explored alternative models, including the double normal and double log-normal distributions. The choice of fit was based on the AD test results and reduced $\chi^2$ values, ensuring that the best possible model was selected for each distribution.

Our results show that in most cases, the flux and index distributions are well-described by either log-normal or double log-normal models. However, for some sources, particularly in the index data,  normal distribution was observed. The mathematical functions used for fitting these distributions (on logarithmic data) are outlined below:

1. Normal Distribution:

\begin{equation}
\label{eq:nor}
N(x) = \frac{10^x \log(10)}{\sqrt{2\pi}\sigma_n}e^{-(10^x-\mu_n)^2/2\sigma_n^2}
\end{equation}

Where $\mu_n$ is the mean and $\sigma_n$ is the standard deviation of the distribution.

2. Log-Normal Distribution:

\begin{equation}
\label{eq:ln}
L(x) = \frac{1}{\sqrt{2\pi}\sigma_l}e^{-(x-\mu_l)^2/2\sigma_l^2}
\end{equation}

Where $\mu_l$ and $\sigma_l$ are the mean and standard deviation of the logarithmic distribution.

and

3. Double Log-Normal Distribution:

\begin{equation}\label{eq:dpdf}
 \rm D(x) = \frac{a}{\sqrt{2\pi \sigma_1^2}} e^\frac{-(x-\mu_1)^2}{2\sigma_1^2} \
       + \frac{(1-a)}{\sqrt{2\pi \sigma_2^2}} e^\frac{-(x-\mu_2)^2}{2\sigma_2^2} 
\end{equation}

where, $\rm a$ is the normalization fraction, $\rm \mu_1$ and $\rm \mu_2$ are the centroids of the distribution with widths $\rm \sigma_1$ and $\rm \sigma_2$ respectively of the logarithmic distribution.

While fitting the above profiles we have kept all the parameters free. In addition to the fitting procedures outlined above, histograms of the flux and index values were constructed to assess how well the distributions adhered to the fitted profiles. These histograms were constructed with an equal number of points per bin, while varying the bin widths to ensure that no bin had zero counts. The results of the best fitting profiles (Figure \ref{fig:fi3day} for 3-day, Figure \ref{fig:fi7day} for 7-day and Figure \ref{fig:fi30day} for 30-day), including the corresponding AD test values and reduced $\chi^2$ values for each source in case of flux and index are summarized in Table \ref{table:flux_par} and Table \ref{table:index_par} respectively.

\subsubsection{Flux and Index Distribution Analysis}

The flux distributions for the VHE FSRQs reveal diverse behaviors across the three different time bins (3-day, 7-day, and 30-day). For seven sources (4FGL J0221.1+3556, 4FGL J0348.5-2749, 4FGL J0904.9-5734, 4FGL J1224.9+2122, 4FGL J1310.5+3221, 4FGL J1443.9+2501, 4FGL J1512.8-0906), the flux distributions in all three bins exhibit a double log-normal profile. This behavior aligns with findings in other studies, where double log-normal distributions are often interpreted as an indication of distinct emission regions contributing to the observed variability \citep{2016ApJ...822L..13K,2018RAA....18..141S}.
Three other sources (4FGL J0739.2+0137, 4FGL J1159.5+2914, 4FGL J1422.3+3223) showed double log-normal behavior in the 3-day and 7-day bins but changed to a single log-normal profile in the 30-day bin. This change could be due to the averaging effects of longer time bins, which may smooth out the finer variability features detected in shorter bins. Similarly, one source (4FGL J1256.1-0547) displayed a log-normal distribution in the 7-day and 30-day bins but showed a double log-normal profile in the 3-day bin, indicating a time-dependent structural change in the emission pattern (See top right panels of Figure \ref{fig:fi3day} for 3-day, Figure \ref{fig:fi7day} for 7-day and Figure \ref{fig:fi30day} for 30-day).

For the spectral index distributions, two sources (4FGL J1224.9+2122 and 4FGL J1512.8-0906) exhibited a double log-normal profile consistently across all three bins. Six sources (4FGL J0348.5-2749, 4FGL J0739.2+0137, 4FGL J0904.9-5734, 4FGL J1159.5+2914, 4FGL J1256.1-0547, and 4FGL J1443.9+2501) followed double log-normal distributions in the 3-day and 7-day bins, but the distributions became log-normal in the 30-day bin. This transition to a simpler log-normal profile at longer timescales might reflect the dominance of fewer, larger-scale processes at work, such as jet instabilities that reduce over short-term fluctuations \citep{2001A&A...377..396R}. Two sources (4FGL J0221.1+3556 and 4FGL J1422.3+3223) showed a double log-normal distribution in the 3-day and 7-day bins, while a normal distribution was observed in the 30-day bin. Another source (4FGL J1310.5+3221) showed a double log-normal distribution in the 3-day bin, a log-normal distribution in the 7-day bin, and a normal distribution in the 30-day bin. This transition suggests that, as bin sizes increase, variability from smaller components diminishes, potentially leaving only a large-scale, more uniform process governing the emission (See bottom panels of Figure \ref{fig:fi3day} for 3-day, Figure \ref{fig:fi7day} for 7-day and Figure \ref{fig:fi30day} for 30-day).

An important observation from this analysis is that for all the sources, a double log-normal distribution profile dominates in finer binning (3-day and mostly in 7-day) for both flux and index values. This profile is often interpreted as evidence of complex processes, such as magnetic reconnection or turbulence in the jet, which can produce bursts of emission over short timescales \citep{2014ApJ...780...87M}. As the time bins increase, the flux and index values show a tendency to simplify, often converging toward a single log-normal or even normal distribution. This pattern, as also discussed in \citet{2020MNRAS.491.1934K} implies that larger time bins smooth out smaller-scale variations, leaving only the underlying long-term behavior in view.


\begin{figure*}
\centering
\includegraphics[scale=0.48,angle=0]{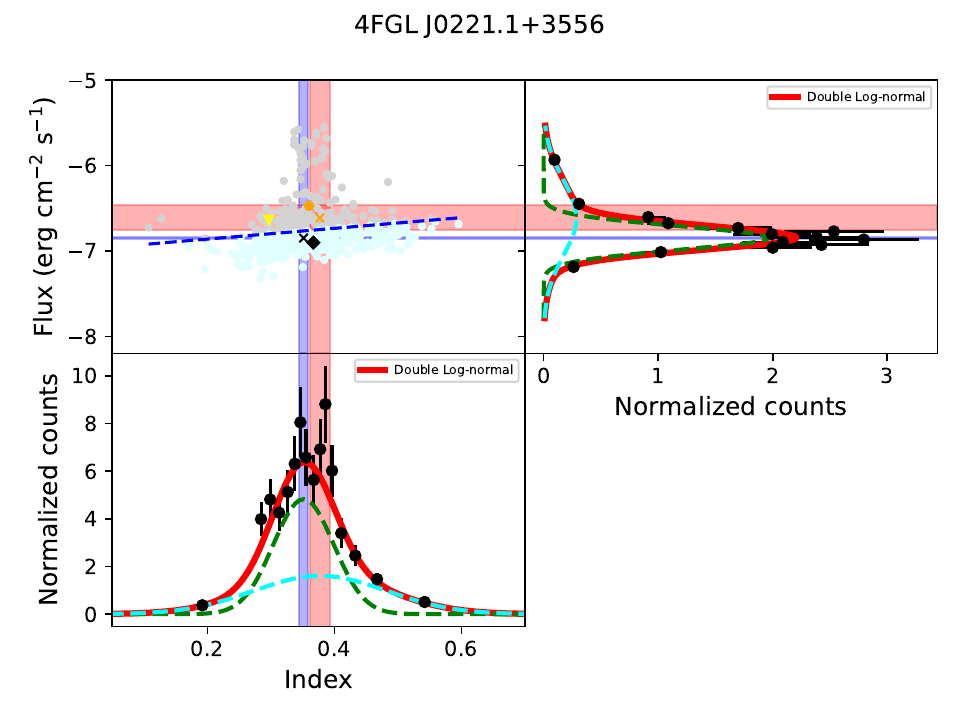}
\includegraphics[scale=0.48,angle=0]{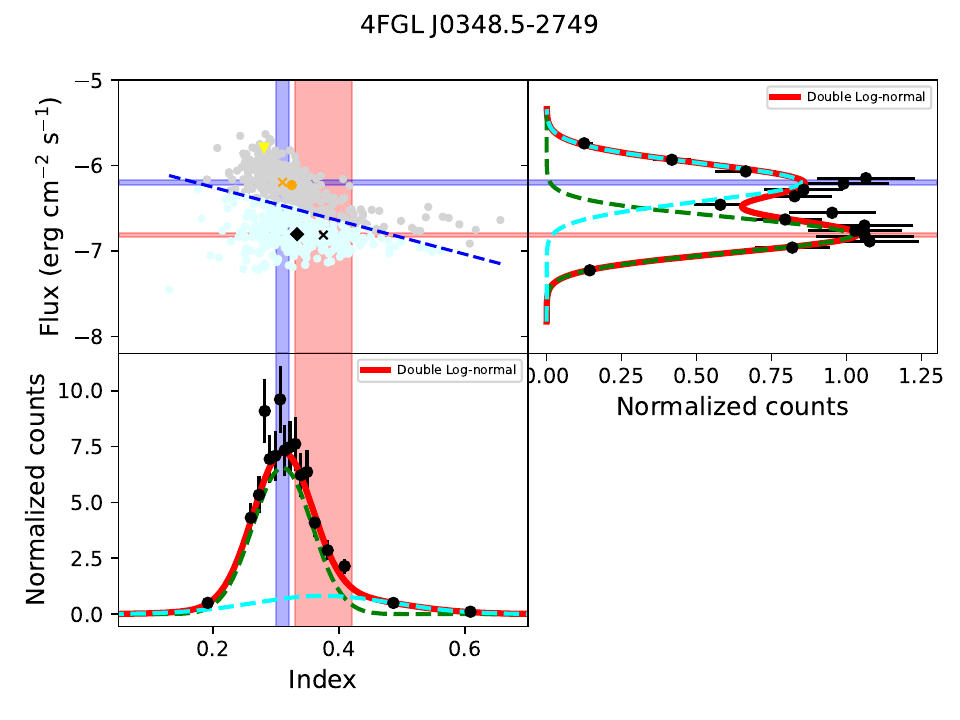}
\includegraphics[scale=0.48,angle=0]{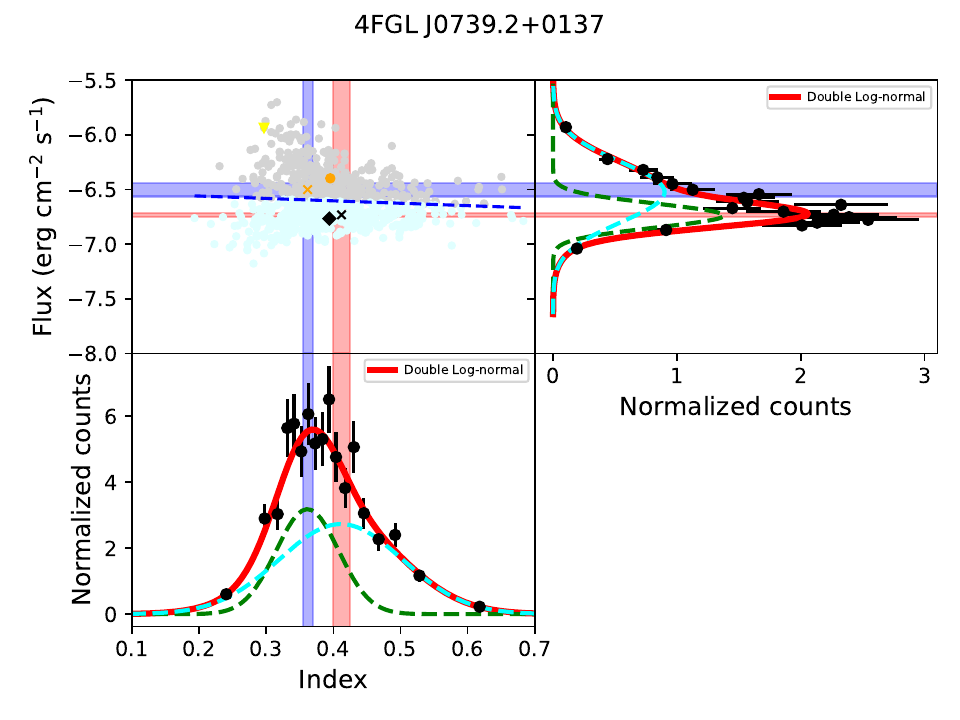}
\includegraphics[scale=0.48,angle=0]{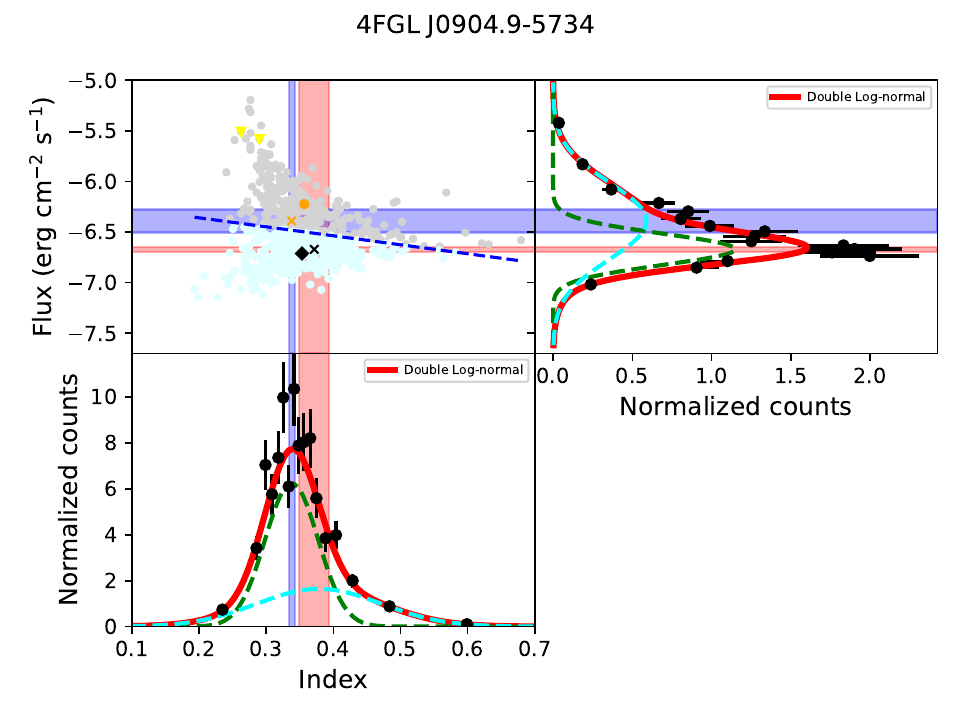}
\includegraphics[scale=0.48,angle=0]{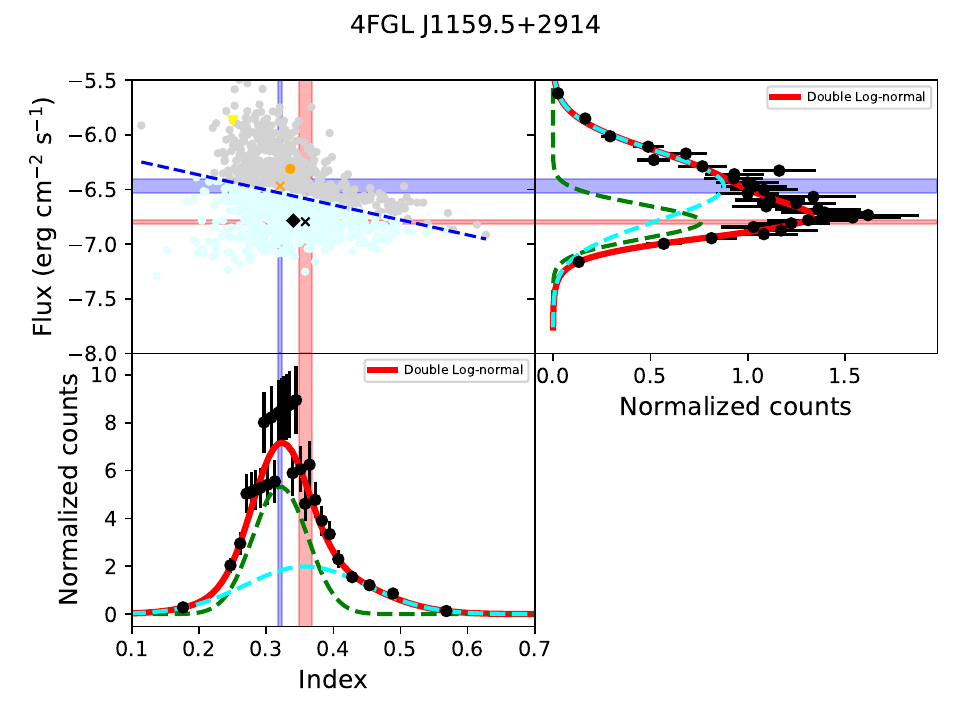}
\includegraphics[scale=0.48,angle=0]{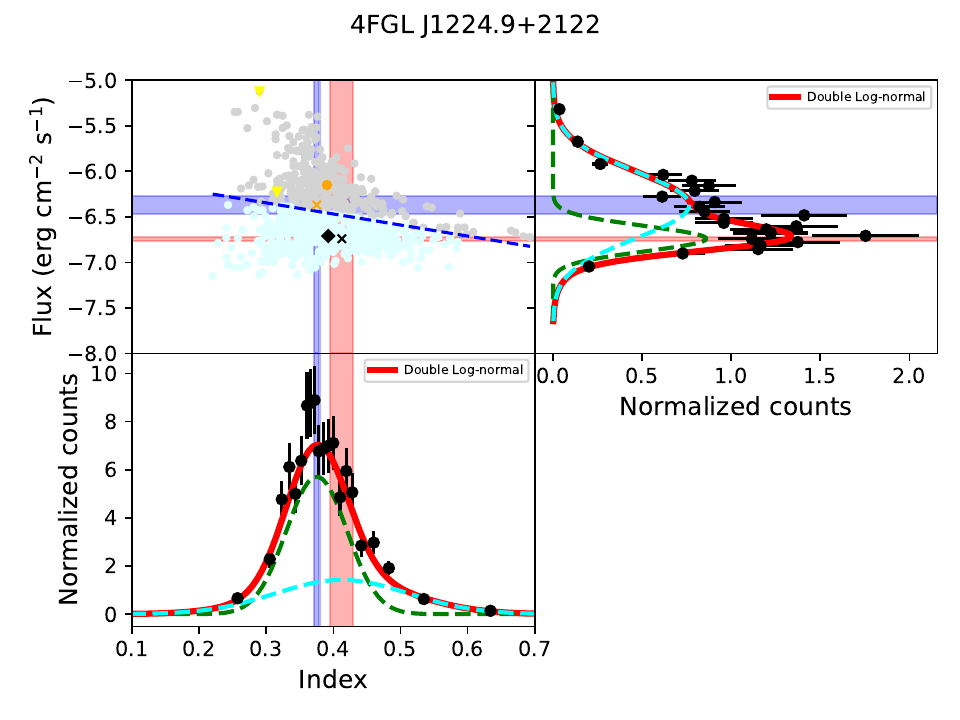}
\hspace*{\fill}
\includegraphics[scale=1,angle=0]{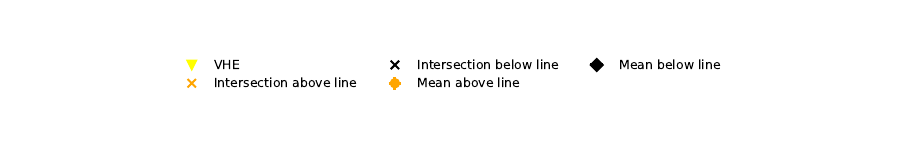}\\

\caption{Multiplot showing the flux/index distribution of 11 VHE sources in 3-day binning. The top left panel displays a scatter plot of the logarithm of flux versus index, along with the best-fit line (dotted blue). Grey and light cyan points are above and below the best-fit line, respectively. The top right panel shows the histogram of the logarithmic flux distribution. The bottom panel shows the histogram of the logarithmic index distribution. The solid red curve (right-hand top panel and bottom panel) represents the best-fitting function, with dotted green and cyan curves showing the individual components in case of double lognormal fit (For normal and lognormal fits, the red curve is replaced by blue and yellow curves, respectively, as indicated in the legend). Vertical and horizontal bands indicate the $1\sigma$ error range for the centroid/centroids of the best fitted function to the index and flux distributions, respectively. The cross symbols in the top left panel indicate the intersection points of the individual centroid/centroids of the distribution, with the yellow inverted triangle representing the corresponding flux and index values at the time the source was detected in VHE. Orange and black filled circles represent the mean values for data points above and below the best-fit line, respectively. The legend for the top left panels is shown at the bottom of the figure.}
\label{fig:fi3day}
\end{figure*}

\begin{figure*}
\setcounter{figure}{1}  
\centering
\includegraphics[scale=0.48,angle=0]{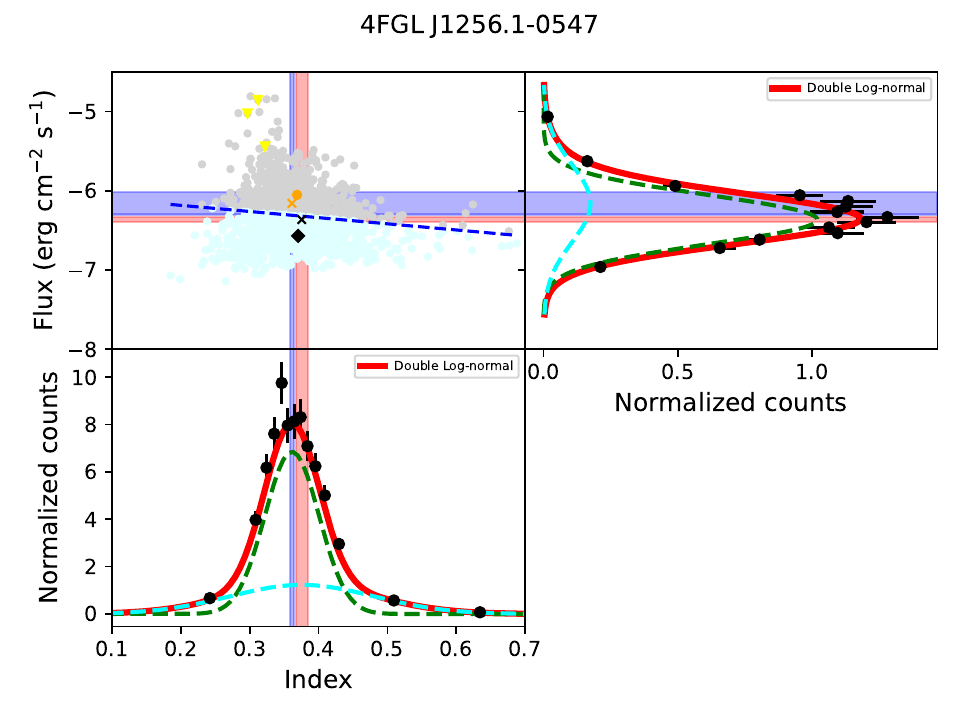}
\includegraphics[scale=0.48,angle=0]{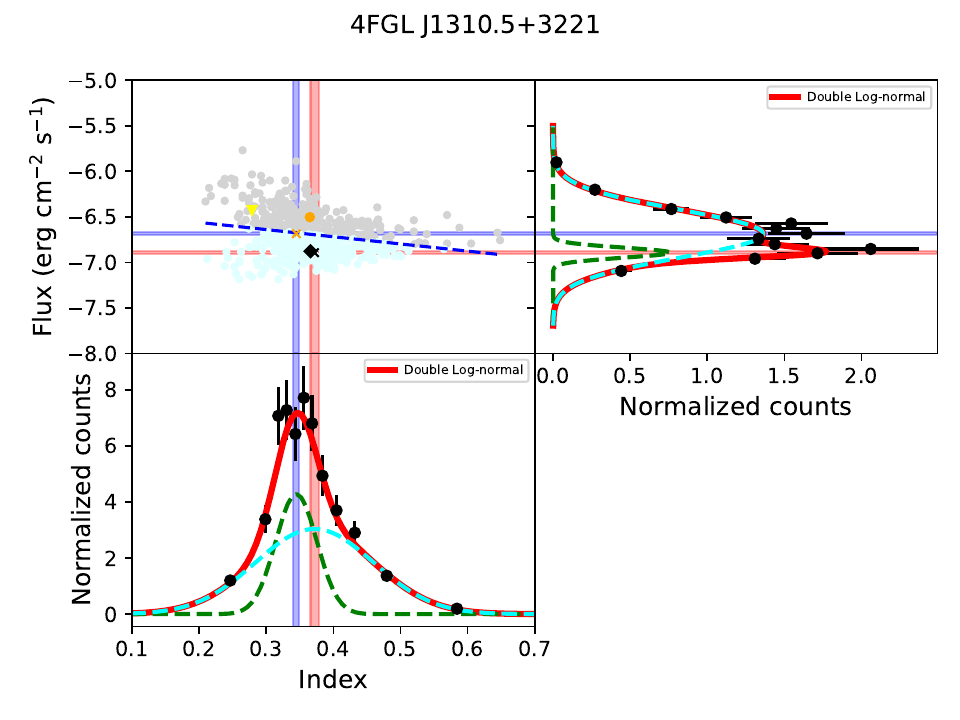}
\includegraphics[scale=0.48,angle=0]{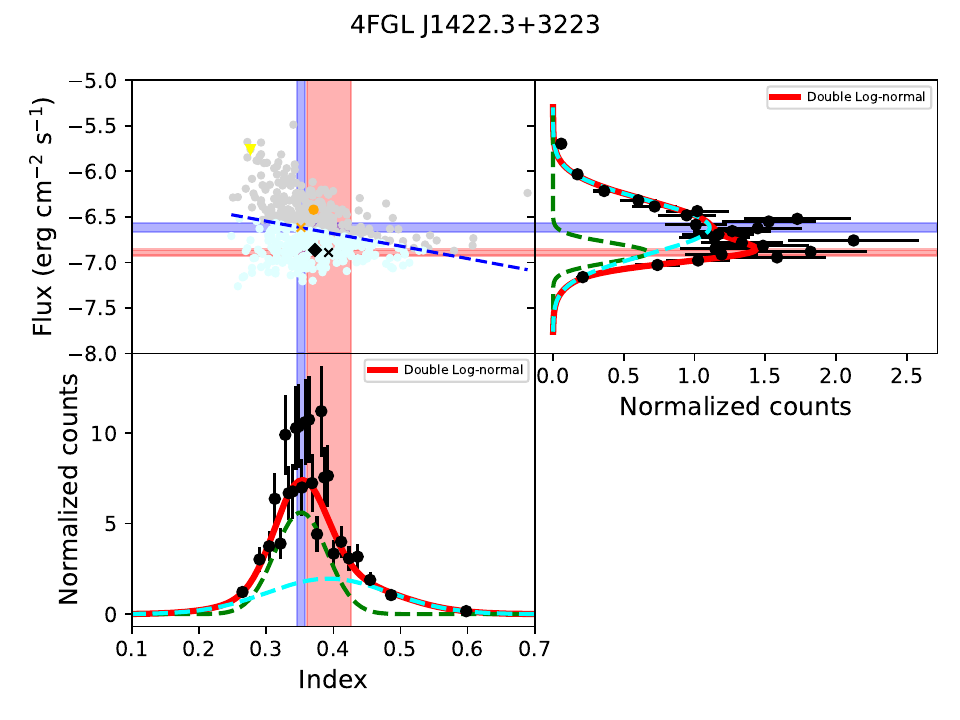}
\includegraphics[scale=0.48,angle=0]{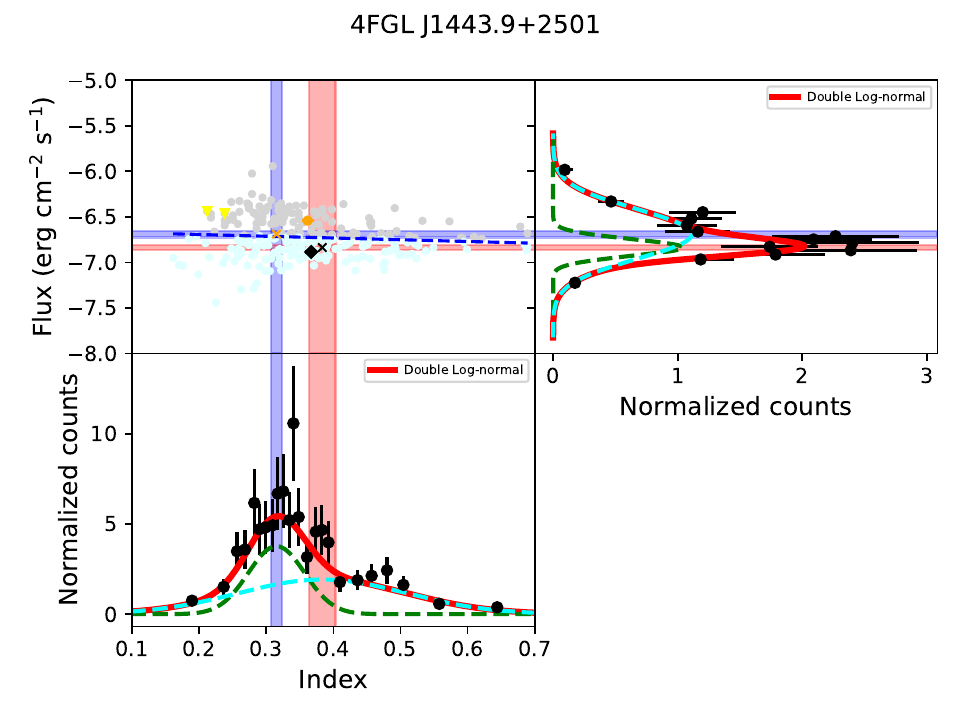}
\includegraphics[scale=0.48,angle=0]{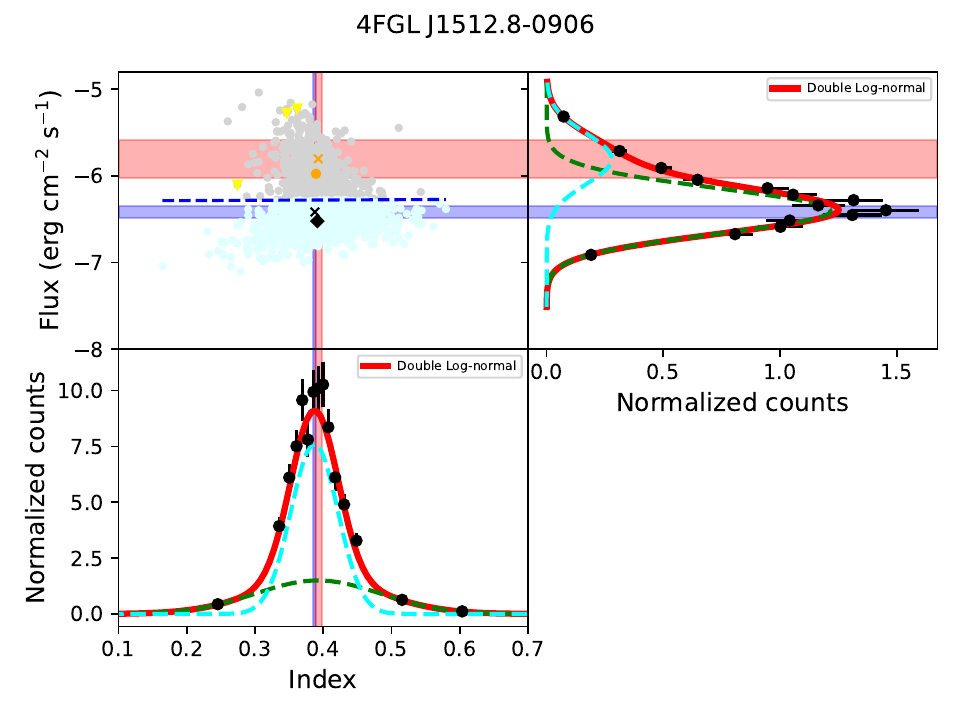}
\hspace*{\fill}
\includegraphics[scale=1,angle=0]{figure/legend.pdf}\\
\caption{(Continued)}
\end{figure*}


\begin{figure*}
\centering
\includegraphics[scale=0.48,angle=0]{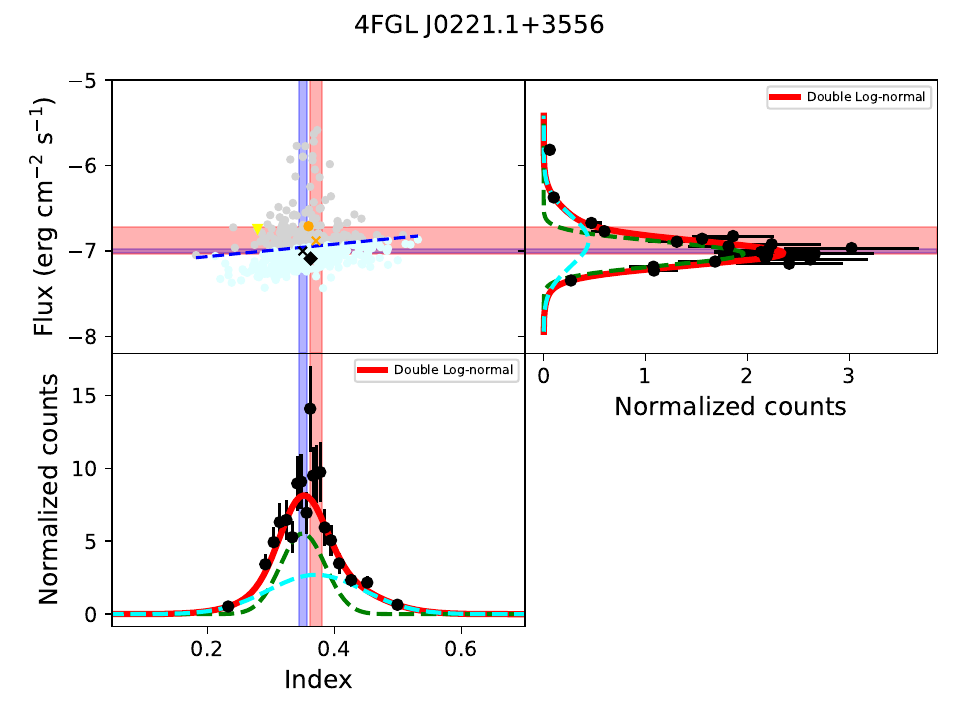}
\includegraphics[scale=0.48,angle=0]{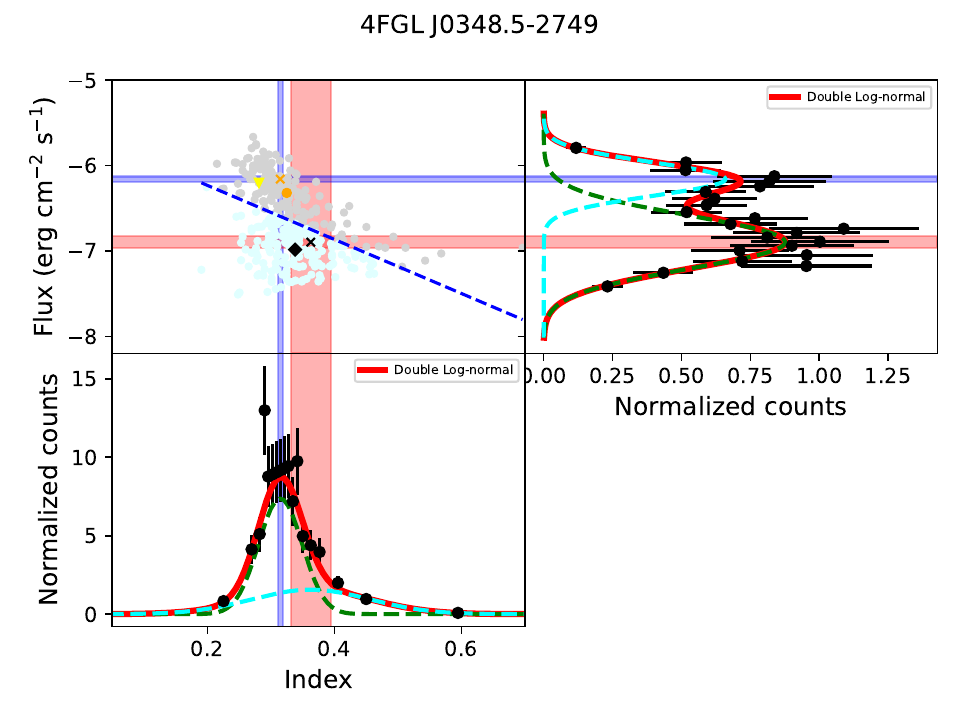}
\includegraphics[scale=0.48,angle=0]{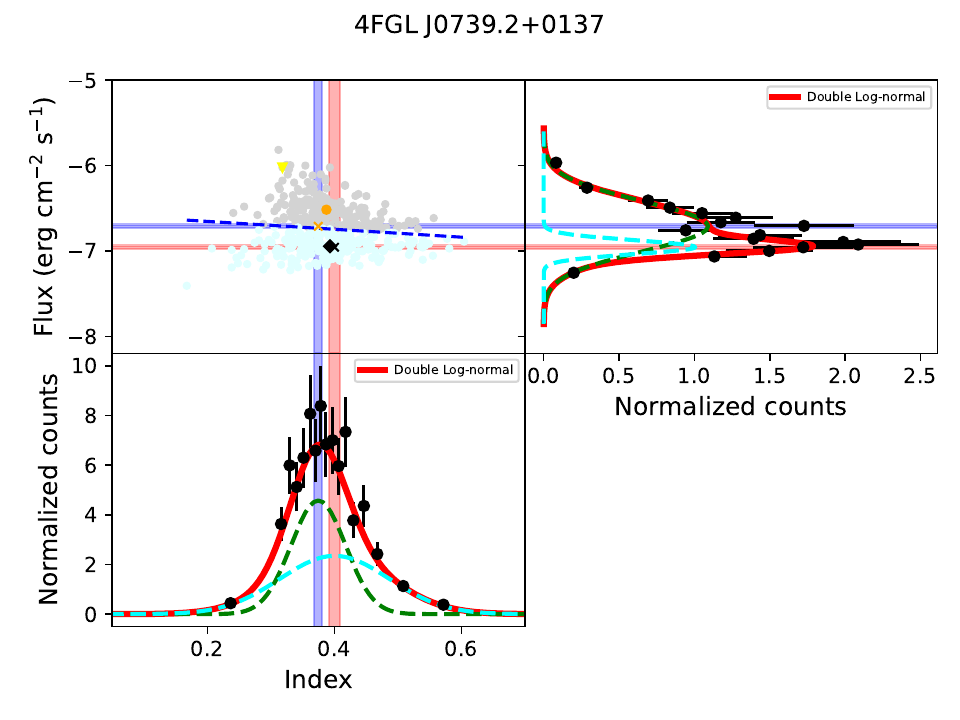}
\includegraphics[scale=0.48,angle=0]{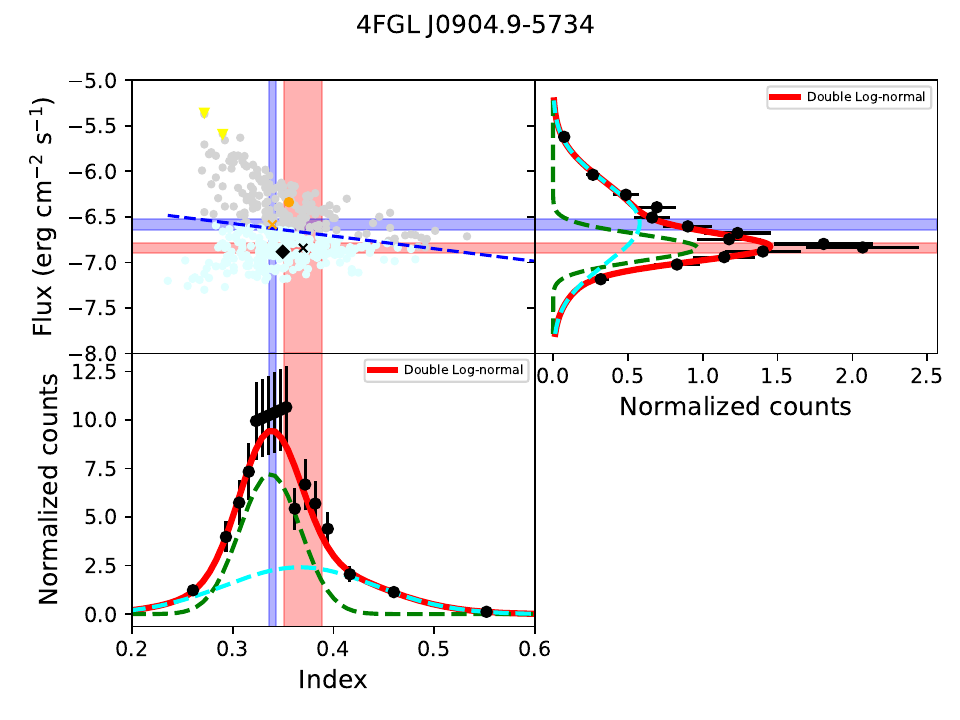}
\includegraphics[scale=0.48,angle=0]{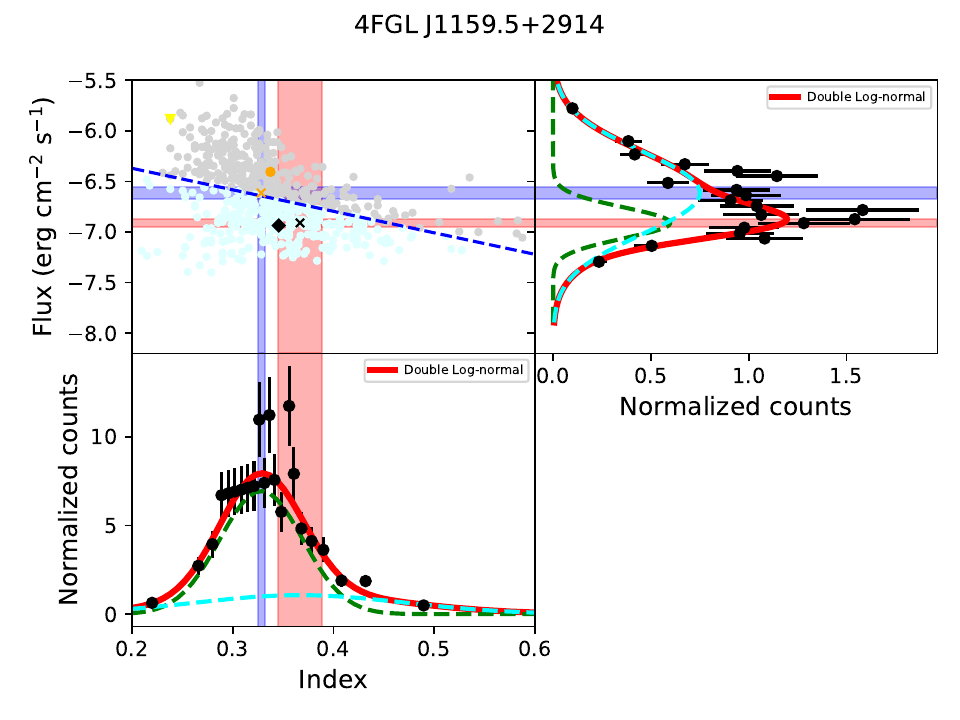}
\includegraphics[scale=0.48,angle=0]{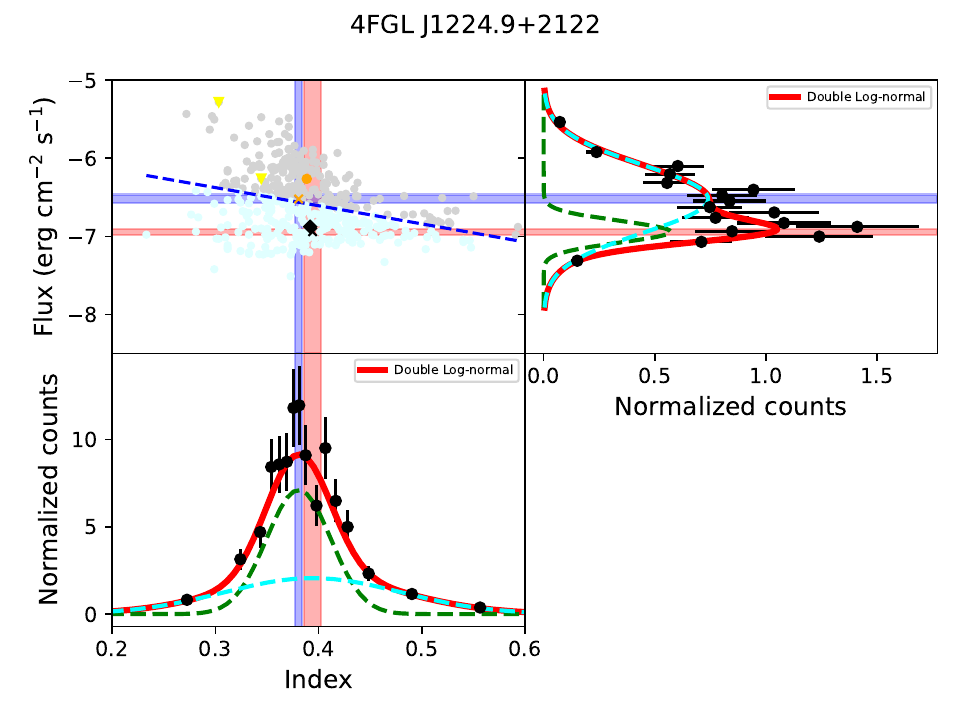}
\hspace*{\fill}
\includegraphics[scale=1,angle=0]{figure/legend.pdf}\\
\caption{Multiplot showing the flux/index distribution of 11 VHE sources in 7-day binning. All other details are the same as in Figure \ref{fig:fi3day}.}
\label{fig:fi7day}
\end{figure*}

\begin{figure*}
\setcounter{figure}{2}  
\centering
\includegraphics[scale=0.48,angle=0]{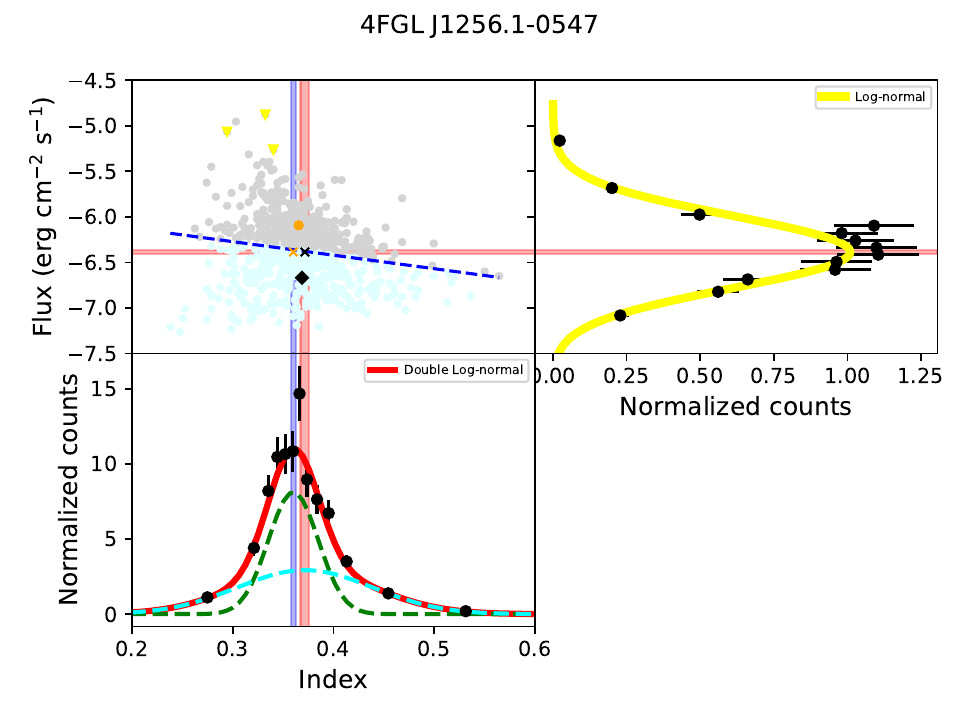}
\includegraphics[scale=0.48,angle=0]{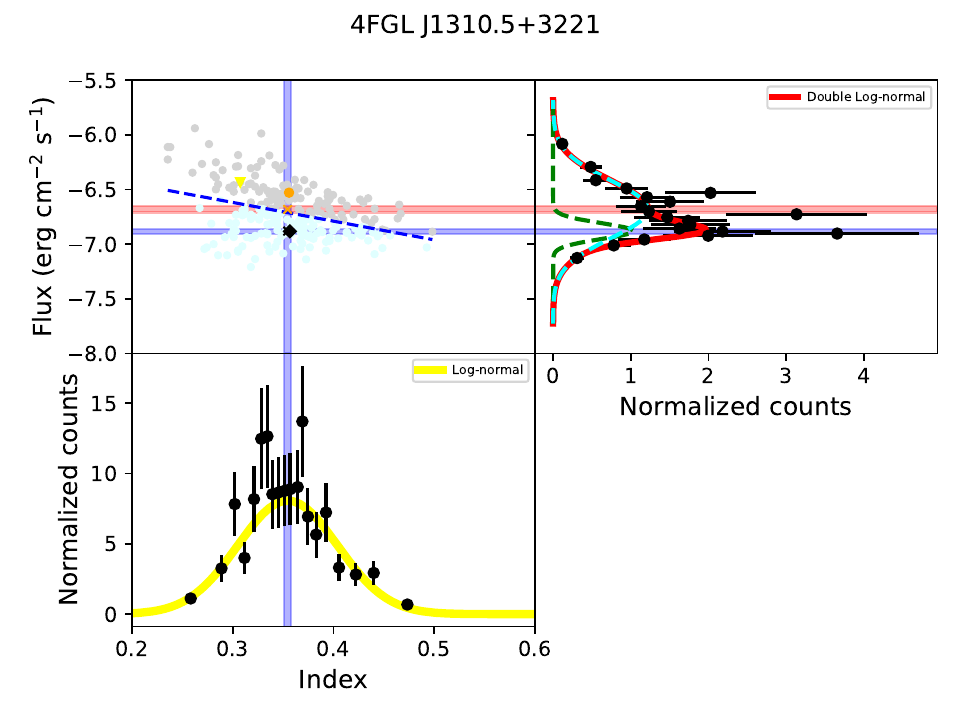}
\includegraphics[scale=0.48,angle=0]{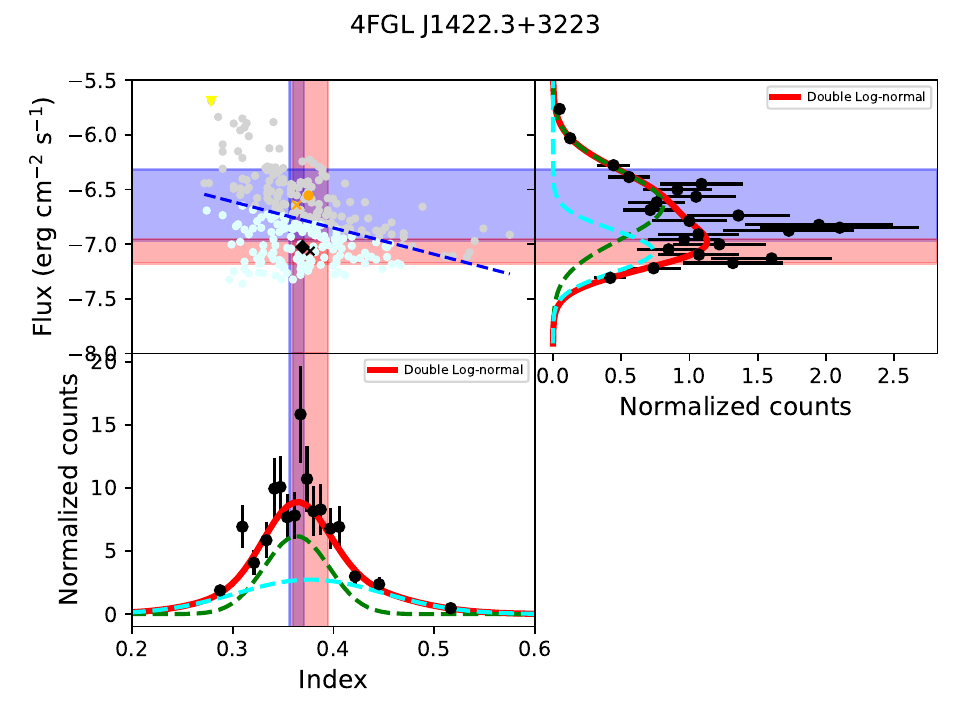}
\includegraphics[scale=0.48,angle=0]{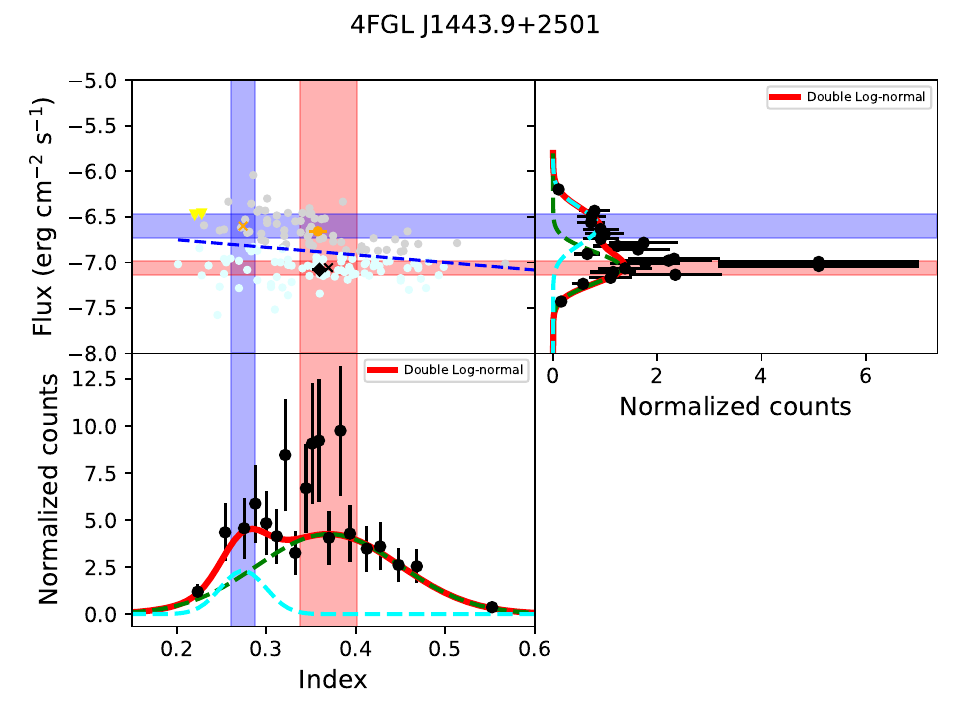}
\includegraphics[scale=0.48,angle=0]{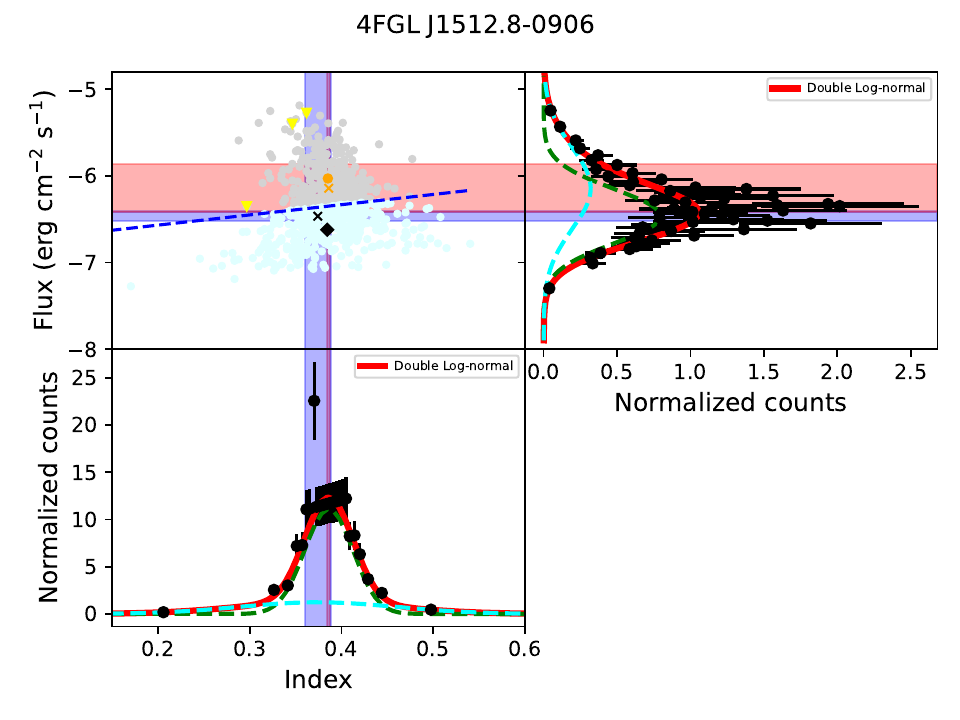}
\hspace*{\fill}
\includegraphics[scale=1,angle=0]{figure/legend.pdf}\\
\caption{(Continued)}
\end{figure*}


\begin{figure*}
\centering
\includegraphics[scale=0.48,angle=0]{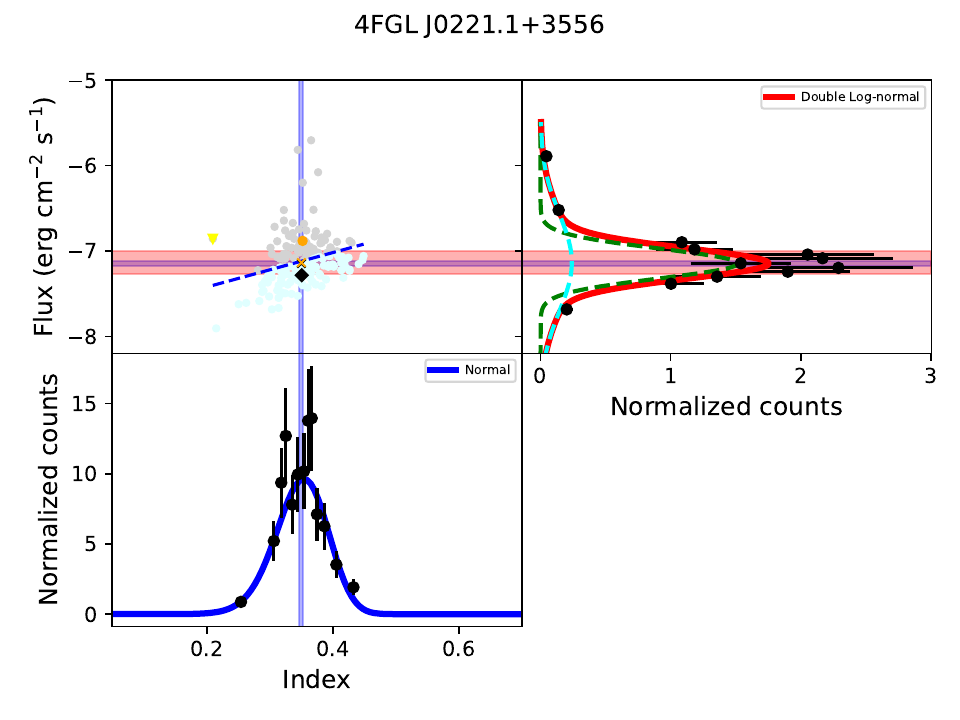}
\includegraphics[scale=0.48,angle=0]{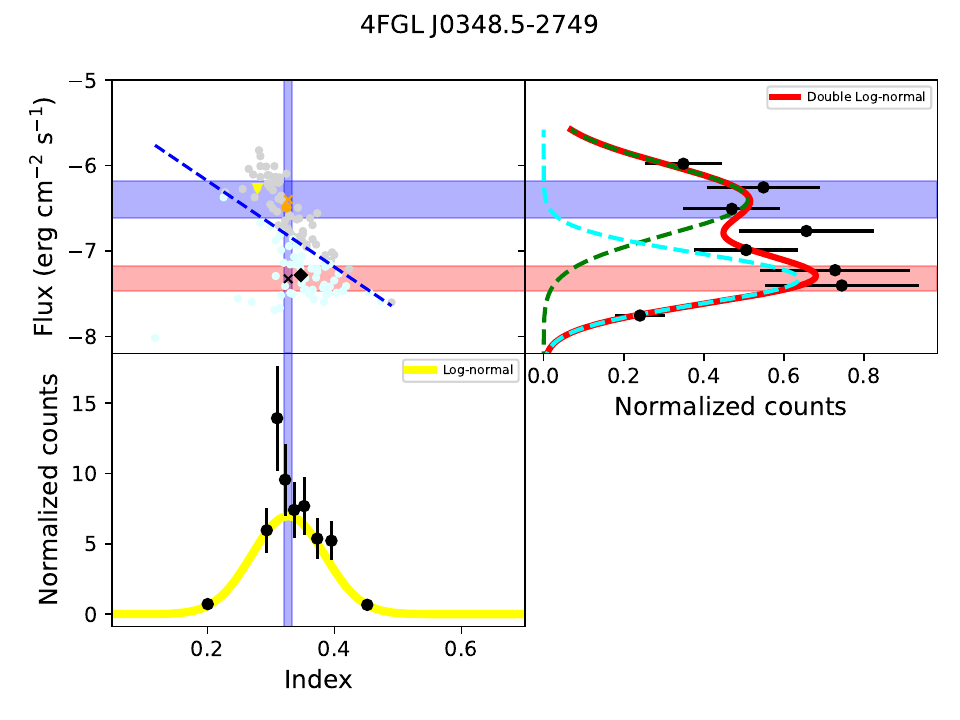}
\includegraphics[scale=0.48,angle=0]{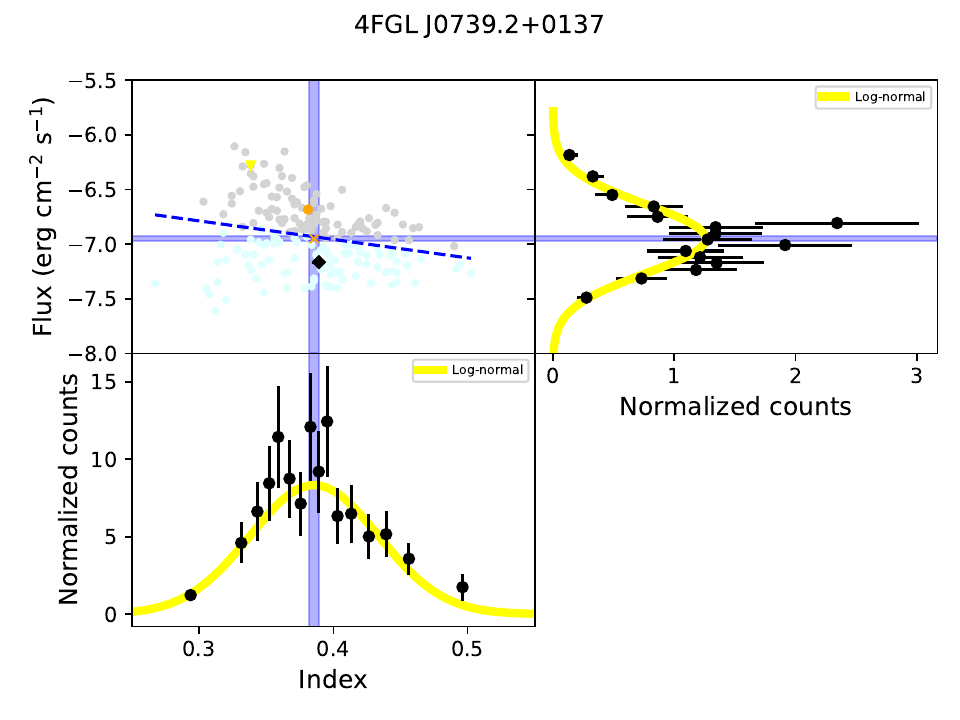}
\includegraphics[scale=0.48,angle=0]{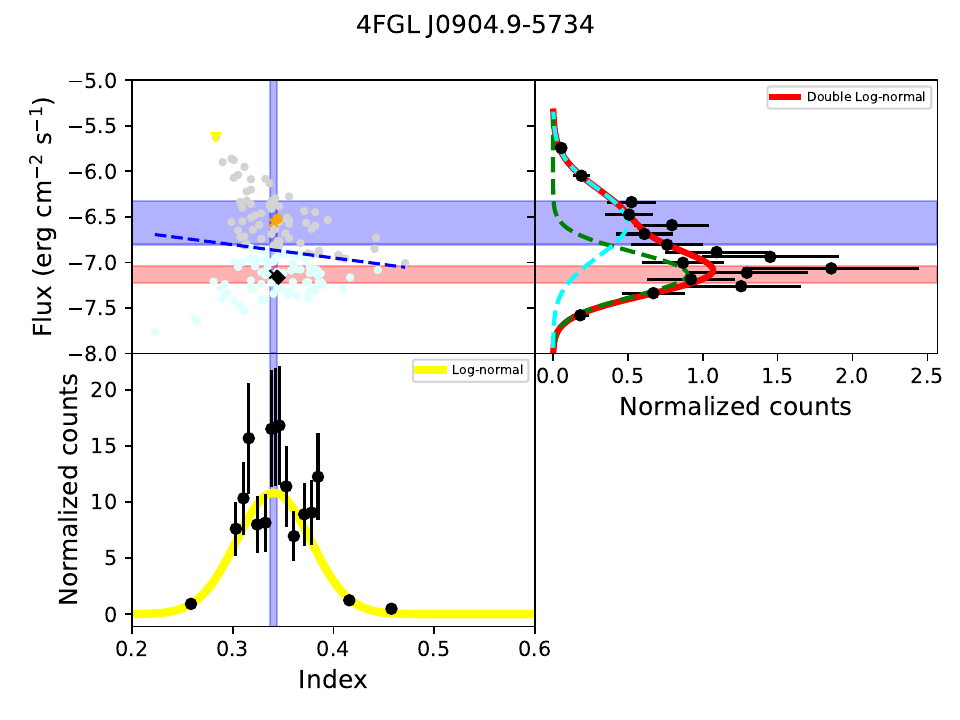}
\includegraphics[scale=0.48,angle=0]{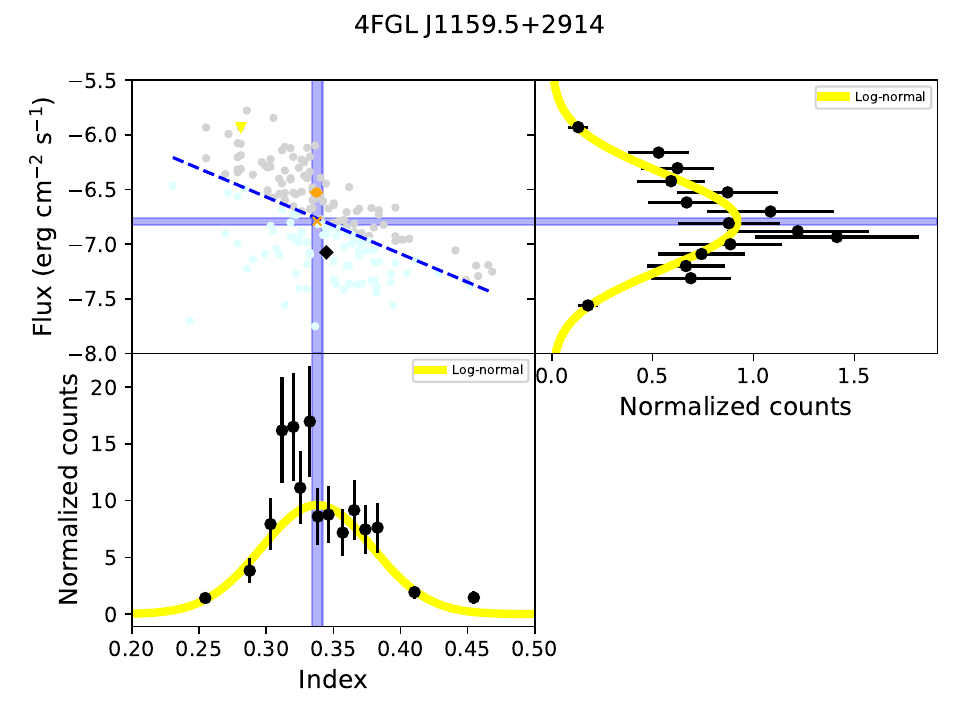}
\includegraphics[scale=0.48,angle=0]{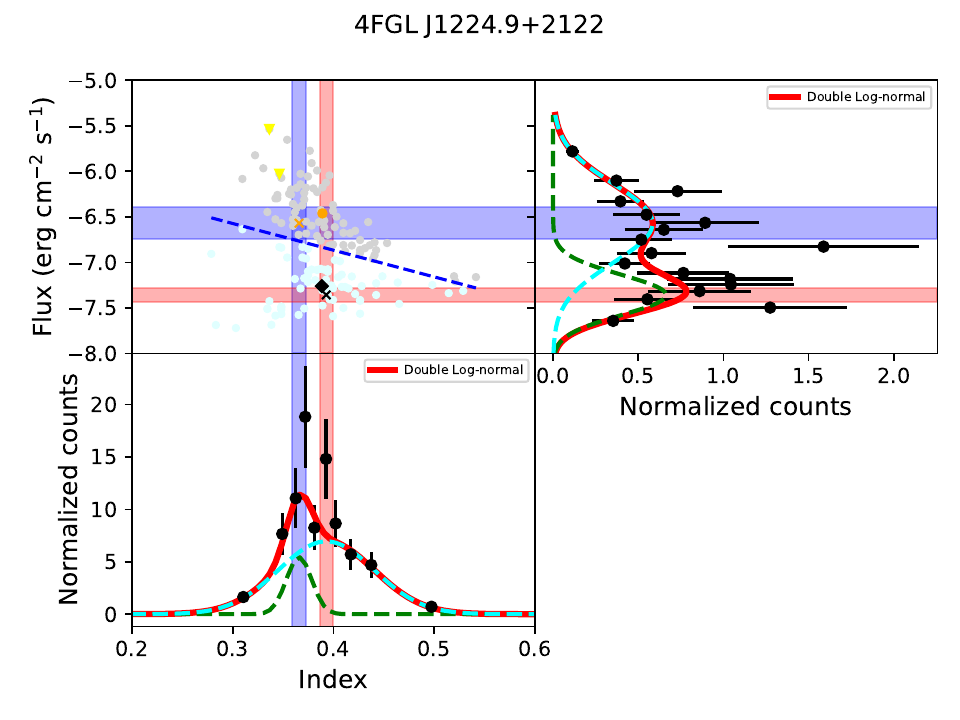}
\hspace*{\fill}
\includegraphics[scale=1,angle=0]{figure/legend.pdf}\\
\caption{Multiplot showing the flux/index distribution of 11 VHE sources in 30-day binning. All other details are the same as in Figure \ref{fig:fi3day}.}
\label{fig:fi30day}
\end{figure*}

\begin{figure*}
\setcounter{figure}{3}  
\centering
\includegraphics[scale=0.48,angle=0]{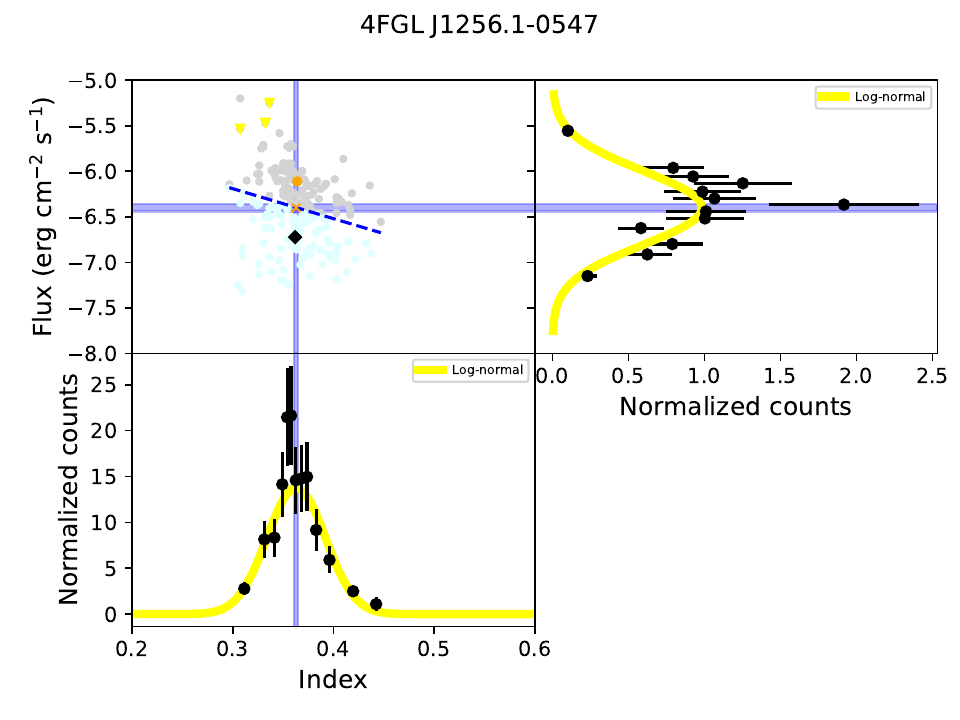}
\includegraphics[scale=0.48,angle=0]{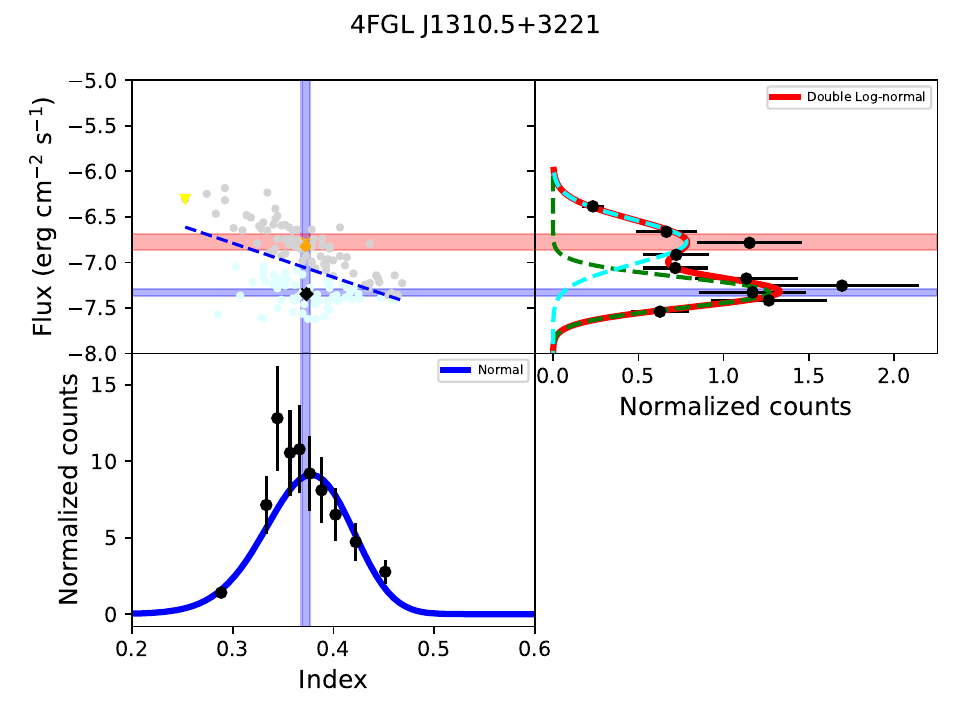}
\includegraphics[scale=0.48,angle=0]{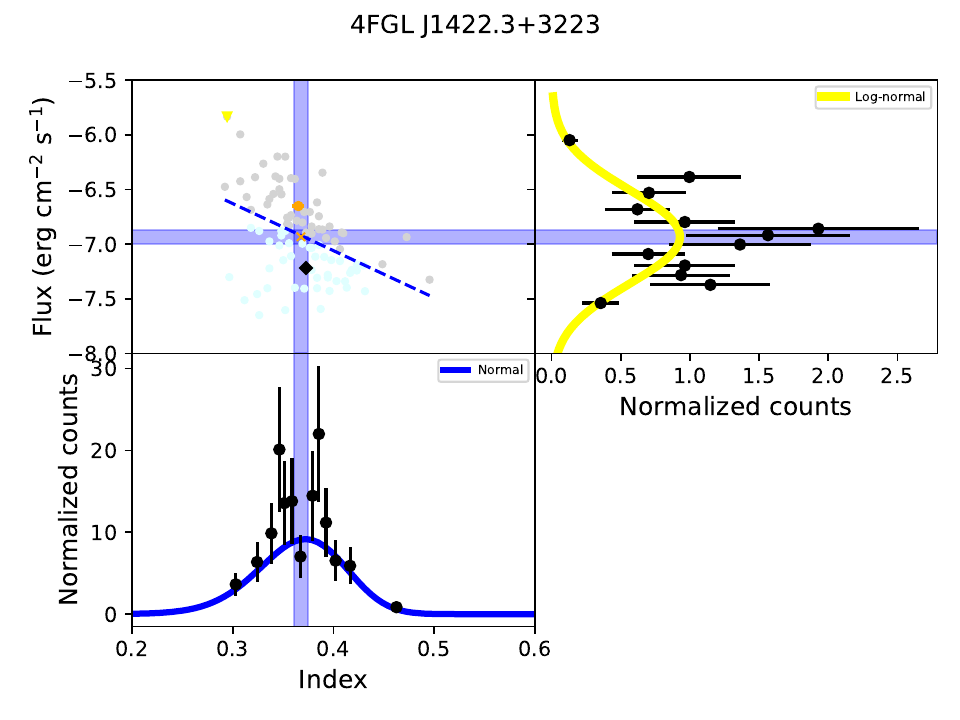}
\includegraphics[scale=0.48,angle=0]{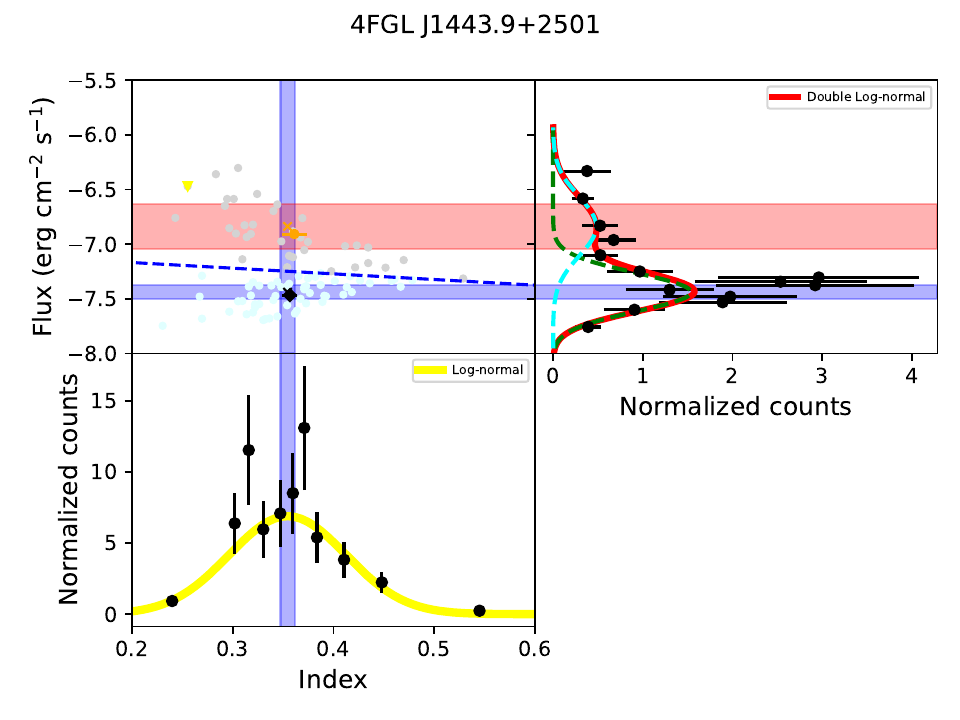}
\includegraphics[scale=0.48,angle=0]{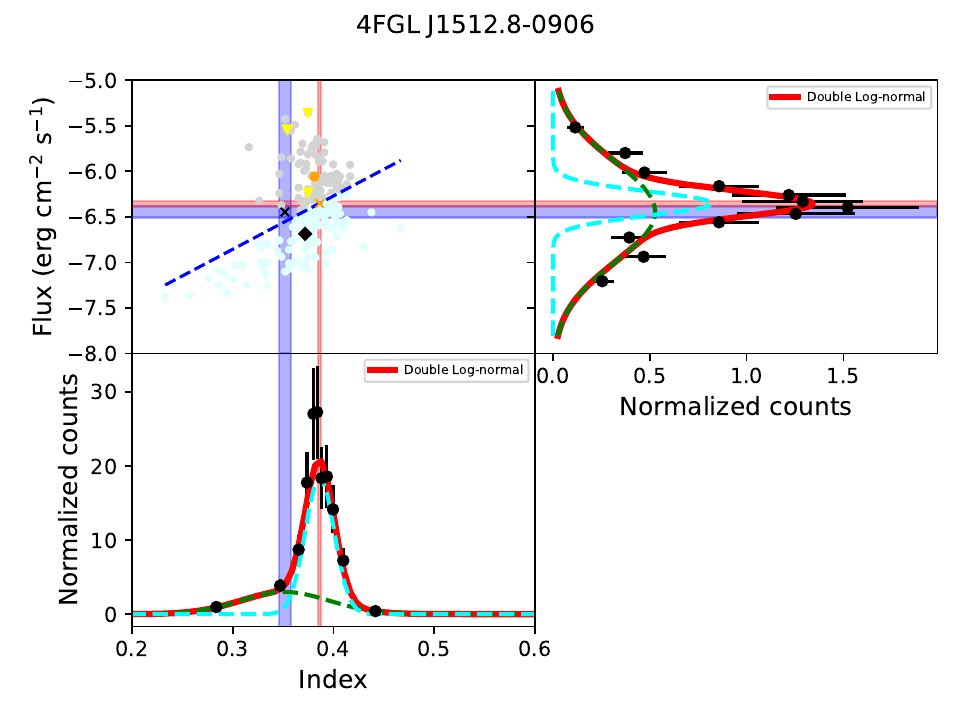}
\hspace*{\fill}
\includegraphics[scale=1,angle=0]{figure/legend.pdf}\\
\caption{(Continued)}
\end{figure*}


\begin{table*}
\caption{Best-fitting parameter values of the function fitted to the flux histograms. Column 1: Source Name; Column 2: Corresponding time bin; Column 3: Fitted function; Column 4: Normalization for the double lognormal fit; Column 5: Best fit values of $\sigma_n$/ $\sigma_l$/ $\sigma_1$; Column 6: Best fit values of $\mu_n$/ $\mu_l$/ $\mu_1$; Column 7: Best fit values of $\sigma_2$; Column 8: Best fit values of $\mu_2$; Column 9: Corresponding reduced $\chi^2$; Column 10: Anderson-Darling (AD) test statistic.
}
\label{table:flux_par}
\centering
\resizebox{\textwidth}{!}{
\begin{tabular}{lllccccccc}
\hline
\textbf{Source} & \textbf{Time Bin} & \textbf{Distribution} & \textbf{a} & \textbf{$\sigma_n$/ $\sigma_l$/ $\sigma_1$} & \textbf{$\mu_n$/ $\mu_l$/ $\mu_1$} & \textbf{$\sigma_2$} & \textbf{$\mu_2$} & \textbf{Reduced Chi-Squared} & \textbf{AD-Test} \\ \hline \hline
4FGL J0221.1+3556 & 1-month & Double Log-normal & $0.659 \pm 0.091$ & $0.174 \pm 0.034$ & $-7.147 \pm 0.029$ & $0.565 \pm 0.192$ & $-7.136 \pm 0.133$ & 1.251 & 2.764 \\ 
& 7-day & Double Log-normal & $0.672 \pm 0.268$ & $0.136 \pm 0.031$ & $-7.032 \pm 0.022$ & $0.297 \pm 0.067$ & $-6.881 \pm 0.158$ & 1.762 & 16.812 \\ 
& 3-day & Double Log-normal & $0.683 \pm 0.108$ & $0.139 \pm 0.018$ & $-6.848 \pm 0.014$ & $0.444 \pm 0.080$ & $-6.613 \pm 0.146$ & 1.292 & 24.716 \\ 
4FGL J0348.5-2749 & 1-month & Double Log-normal & $0.522 \pm 0.249$ & $0.408 \pm 0.224$ & $-6.397 \pm 0.217$ & $0.299 \pm 0.095$ & $-7.326 \pm 0.147$ & 1.245 & 1.531 \\ 
& 7-day & Double Log-normal & $0.679 \pm 0.044$ & $0.309 \pm 0.028$ & $-6.897 \pm 0.031$ & $0.194 \pm 0.020$ & $-6.158 \pm 0.071$ & 0.508 & 3.008 \\ 
& 3-day & Double Log-normal & $0.526 \pm 0.045$ & $0.206 \pm 0.016$ & $-6.814 \pm 0.027$ & $0.223 \pm 0.021$ & $-6.196 \pm 0.030$ & 0.926 & 5.696 \\ 
4FGL J0739.2+0137 & 1-month & Log-normal & - & $0.317 \pm 0.020$ & $-6.951 \pm 0.022$ & - & - & 0.740 & 0.559 \\ 
& 7-day & Double Log-normal & $0.798 \pm 0.053$ & $0.292 \pm 0.015$ & $-6.711 \pm 0.021$ & $0.080 \pm 0.021$ & $-6.955 \pm 0.019$ & 1.121 & 2.956 \\ 
& 3-day & Double Log-normal & $0.428 \pm 0.138$ & $0.121 \pm 0.028$ & $-6.732 \pm 0.018$ & $0.284 \pm 0.023$ & $-6.501 \pm 0.063$ & 1.754 & 9.313 \\
4FGL J0904.9-5734 & 1-month & Double Log-normal & $0.538 \pm 0.370$ & $0.239 \pm 0.054$ & $-7.131 \pm 0.092$ & $0.369 \pm 0.167$ & $-6.566 \pm 0.238$ & 0.659 & 1.364 \\ 
& 7-day &  Double Log-normal & $0.314 \pm 0.082$ & $0.133 \pm 0.028$ & $-6.843 \pm 0.052$ & $0.449 \pm 0.033$ & $-6.586 \pm 0.059$ & 0.976 & 6.189 \\ 
& 3-day & Double Log-normal & $0.443 \pm 0.167$ & $0.154 \pm 0.035$ & $-6.671 \pm 0.022$ & $0.376 \pm 0.042$ & $-6.391 \pm 0.114$ & 1.447 & 9.311 \\ 
4FGL J1159.5+2914 & 1-month & Log-normal & - & $0.434 \pm 0.025$ & $-6.790 \pm 0.031$ & - & - & 0.690 & 0.578 \\
& 7-day & Double Log-normal & $0.230 \pm 0.112$ & $0.153 \pm 0.050$ & $-6.910 \pm 0.038$ & $0.410 \pm 0.028$ & $-6.613 \pm 0.062$ & 1.452 & 3.441 \\ 
& 3-day & Double Log-normal & $0.297 \pm 0.128$ & $0.156 \pm 0.034$ & $-6.795 \pm 0.017$ & $0.320 \pm 0.022$ & $-6.469 \pm 0.062$ & 0.694 & 5.999 \\ 
4FGL J1224.9+2122 & 1-month & Double Log-normal & $0.367 \pm 0.185$ & $0.221 \pm 0.074$ & $-7.358 \pm 0.080$ & $0.432 \pm 0.112$ & $-6.570 \pm 0.178$ & 1.046 & 1.088 \\ 
& 7-day & Double Log-normal & $0.197 \pm 0.094$ & $0.137 \pm 0.057$ & $-6.941 \pm 0.038$ & $0.433 \pm 0.028$ & $-6.521 \pm 0.057$ & 1.203 & 3.376 \\ 
& 3-day & Double Log-normal & $0.296 \pm 0.147$ & $0.137 \pm 0.041$ & $-6.740 \pm 0.023$ & $0.368 \pm 0.044$ & $-6.370 \pm 0.095$ & 1.578 & 8.096 \\ 
4FGL J1256.1-0547 & 1-month & Log-normal & - & $0.406 \pm 0.028$ & $-6.401 \pm 0.040$ & - & - & 1.445 & 0.800 \\ 
& 7-day & Log-normal & - & $0.395 \pm 0.015$ & $-6.386 \pm 0.018$ & - & - & 1.528 & 1.528 \\ 
& 3-day & Double Log-normal & $0.786 \pm 0.398$ & $0.308 \pm 0.055$ & $-6.364 \pm 0.032$ & $0.489 \pm 0.097$ & $-6.158 \pm 0.141$ & 1.636 & 3.301 \\ 
4FGL J1310.5+3221 & 1-month & Double Log-normal & $0.523 \pm 0.120$ & $0.164 \pm 0.035$ & $-7.332 \pm 0.039$ & $0.243 \pm 0.059$ & $-6.775 \pm 0.085$ & 0.794 & 1.659 \\ 
& 7-day & Double Log-normal & $0.187 \pm 0.072$ & $0.072 \pm 0.024$ & $-6.881 \pm 0.023$ & $0.266 \pm 0.019$ & $-6.682 \pm 0.028$ & 0.979 & 1.851 \\ 
& 3-day & Double Log-normal & $0.105 \pm 0.035$ & $0.055 \pm 0.018$ & $-6.894 \pm 0.017$ & $0.262 \pm 0.011$ & $-6.686 \pm 0.015$ & 0.877 & 2.381 \\ 
4FGL J1422.3+3223 & 1-month & Log-normal & - & $0.430 \pm 0.051$ & $-6.933 \pm 0.063$ & - & - & 1.021 & 0.417 \\ 
& 7-day & Double Log-normal & $0.643 \pm 0.284$ & $0.318 \pm 0.134$ & $-6.641 \pm 0.325$ & $0.194 \pm 0.110$ & $-7.063 \pm 0.113$ & 1.171 & 2.320 \\ 
& 3-day & Double Log-normal & $0.182 \pm 0.119$ & $0.107 \pm 0.050$ & $-6.892 \pm 0.033$ & $0.295 \pm 0.022$ & $-6.617 \pm 0.048$ & 1.367 & 3.943 \\ 
4FGL J1443.9+2501 & 1-month & Double Log-normal & $0.674 \pm 0.240$ & $0.175 \pm 0.043$ & $-7.440 \pm 0.063$ & $0.270 \pm 0.234$ & $-6.838 \pm 0.209$ & 1.134 & 2.216 \\ 
& 7-day & Double Log-normal & $0.571 \pm 0.209$ & $0.174 \pm 0.039$ & $-7.058 \pm 0.076$ & $0.197 \pm 0.065$ & $-6.602 \pm 0.133$ & 1.295 & 1.203 \\ 
& 3-day & Double Log-normal & $0.241 \pm 0.122$ & $0.092 \pm 0.035$ & $-6.837 \pm 0.031$ & $0.272 \pm 0.033$ & $-6.689 \pm 0.037$ & 1.234 & 1.397 \\ 
4FGL J1512.8-0906 & 1-month & Double Log-normal & $0.729 \pm 0.087$ & $0.551 \pm 0.076$ & $-6.446 \pm 0.061$ & $0.132 \pm 0.039$ & $-6.354 \pm 0.032$ & 0.901 & 0.902 \\ 
& 7-day & Double Log-normal & $0.621 \pm 0.304$ & $0.312 \pm 0.063$ & $-6.464 \pm 0.054$ & $0.472 \pm 0.064$ & $-6.141 \pm 0.277$ & 0.893 & 1.394 \\ 
& 3-day & Double Log-normal & $0.795 \pm 0.132$ & $0.261 \pm 0.020$ & $-6.415 \pm 0.065$ & $0.294 \pm 0.103$ & $-5.802 \pm 0.218$ & 0.989 & 10.135 \\ \hline
\end{tabular}
}
\end{table*}

\begin{table*}
\caption{Best-fitting parameter values of the function fitted to the index histograms. Column 1: Source Name; Column 2: Corresponding time bin; Column 3: Fitted function; Column 4: Normalization for the double lognormal fit; Column 5: Best fit values of $\sigma_n$/ $\sigma_l$/ $\sigma_1$; Column 6: Best fit values of $\mu_n$/ $\mu_l$/ $\mu_1$; Column 7: Best fit values of $\sigma_2$; Column 8: Best fit values of $\mu_2$; Column 9: Corresponding reduced $\chi^2$; Column 10: Anderson-Darling (AD) test statistic.
}
\label{table:index_par}
\centering
\resizebox{\textwidth}{!}{
\begin{tabular}{lllccccccc}
\hline 
\textbf{Source} & \textbf{Time Bin} & \textbf{Distribution} & \textbf{a} & \textbf{$\sigma_n$/ $\sigma_l$/ $\sigma_1$} & \textbf{$\mu_n$/ $\mu_l$/ $\mu_1$} & \textbf{$\sigma_2$} & \textbf{$\mu_2$} & \textbf{Reduced $\chi^2$} & \textbf{AD-Test} \\ \hline \hline
4FGL J0221.1+3556 & 1-month & Normal & - & $0.214 \pm 0.012$ & $2.239 \pm 0.016$ & - & - & 0.959 & 0.638 \\ 
& 7-day & Double Log-normal & $0.490 \pm 0.247$ & $0.035 \pm 0.009$ & $0.350 \pm 0.006$ & $0.075 \pm 0.017$ & $0.371 \pm 0.010$ & 1.327 & 1.569 \\ 
& 3-day & Double Log-normal & $0.575 \pm 0.254$ & $0.048 \pm 0.011$ & $0.351 \pm 0.007$ & $0.105 \pm 0.033$ & $0.377 \pm 0.015$ & 1.368 & 1.780 \\ 
4FGL J0348.5-2749 & 1-month & Log-normal & - & $0.057 \pm 0.004$ & $0.327 \pm 0.006$ & - & - & 1.238 & 0.820 \\ 
& 7-day & Double Log-normal & $0.636 \pm 0.096$ & $0.034 \pm 0.005$ & $0.315 \pm 0.004$ & $0.093 \pm 0.009$ & $0.363 \pm 0.031$ & 0.899 & 6.626 \\ 
& 3-day & Double Log-normal & $0.771 \pm 0.108$ & $0.047 \pm 0.005$ & $0.310 \pm 0.008$ & $0.111 \pm 0.018$ & $0.375 \pm 0.045$ & 1.188 & 12.449 \\ 
4FGL J0739.2+0137 & 1-month & Log-normal & - & $0.048 \pm 0.003$ & $0.386 \pm 0.004$ & - & - & 0.721 & 0.412 \\ 
& 7-day & Double Log-normal & $0.493 \pm 0.182$ & $0.043 \pm 0.008$ & $0.374 \pm 0.006$ & $0.086 \pm 0.013$ & $0.400 \pm 0.009$ & 0.943 & 1.944 \\ 
& 3-day & Double Log-normal & $0.382 \pm 0.143$ & $0.047 \pm 0.010$ & $0.362 \pm 0.007$ & $0.093 \pm 0.005$ & $0.412 \pm 0.013$ & 1.099 & 4.106 \\ 
4FGL J0904.9-5734 & 1-month & Log-normal & - & $0.037 \pm 0.002$ & $0.341 \pm 0.003$ & - & - & 1.595 & 0.608 \\ 
& 7-day &  Double Log-normal & $0.617 \pm 0.128$ & $0.033 \pm 0.004$ & $0.339 \pm 0.004$ & $0.076 \pm 0.007$ & $0.370 \pm 0.019$ & 0.843 & 4.421 \\ 
& 3-day & Double Log-normal & $0.619 \pm 0.110$ & $0.039 \pm 0.005$ & $0.338 \pm 0.004$ & $0.092 \pm 0.009$ & $0.371 \pm 0.022$ & 1.589 & 9.380 \\ 
4FGL J1159.5+2914 & 1-month & Log-normal & - & $0.041 \pm 0.003$ & $0.338 \pm 0.004$ & - & - & 1.420 & 0.704 \\ 
& 7-day & Double Log-normal & $0.720 \pm 0.082$ & $0.041 \pm 0.004$ & $0.328 \pm 0.003$ & $0.104 \pm 0.012$ & $0.367 \pm 0.022$ & 1.192 & 6.288 \\ 
& 3-day & Double Log-normal & $0.548 \pm 0.069$ & $0.041 \pm 0.004$ & $0.321 \pm 0.003$ & $0.090 \pm 0.005$ & $0.358 \pm 0.006$ & 1.157 & 10.881 \\ 
4FGL J1224.9+2122 & 1-month & Double Log-normal & $0.172 \pm 0.111$ & $0.013 \pm 0.008$ & $0.366 \pm 0.007$ & $0.048 \pm 0.005$ & $0.393 \pm 0.006$ & 1.873 & 1.696 \\ 
& 7-day & Double Log-normal & $0.556 \pm 0.093$ & $0.031 \pm 0.004$ & $0.381 \pm 0.003$ & $0.086 \pm 0.013$ & $0.394 \pm 0.008$ & 1.103 & 4.476 \\ 
& 3-day & Double Log-normal & $0.647 \pm 0.103$ & $0.045 \pm 0.005$ & $0.375 \pm 0.004$ & $0.099 \pm 0.010$ & $0.413 \pm 0.017$ & 1.314 & 6.368 \\ 
4FGL J1256.1-0547 & 1-month & Log-normal & - & $0.029 \pm 0.002$ & $0.363 \pm 0.002$ & - & - & 0.929 & 0.733 \\ 
& 7-day & Double Log-normal & $0.505 \pm 0.066$ & $0.025 \pm 0.002$ & $0.360 \pm 0.002$ & $0.067 \pm 0.006$ & $0.372 \pm 0.004$ & 1.219 & 4.677 \\ 
& 3-day & Double Log-normal & $0.677 \pm 0.048$ & $0.039 \pm 0.003$ & $0.361 \pm 0.003$ & $0.105 \pm 0.011$ & $0.375 \pm 0.008$ & 1.876 & 11.096 \\ 
4FGL J1310.5+3221 & 1-month & Normal & - & $0.239 \pm 0.019$ & $2.364 \pm 0.023$ & - & - & 1.121 & 0.455 \\ 
& 7-day & Log-normal & - & $0.049 \pm 0.003$ & $0.355 \pm 0.003$ & - & - & 1.042 & 0.597 \\ 
& 3-day & Double Log-normal & $0.330 \pm 0.069$ & $0.031 \pm 0.005$ & $0.345 \pm 0.004$ & $0.088 \pm 0.006$ & $0.372 \pm 0.006$ & 1.004 & 3.181 \\ 
4FGL J1422.3+3223 & 1-month & Normal & - & $0.235 \pm 0.027$ & $2.333 \pm 0.038$ & - & - & 1.098 & 0.518 \\ 
& 7-day & Double Log-normal & $0.498 \pm 0.209$ & $0.032 \pm 0.010$ & $0.363 \pm 0.007$ & $0.073 \pm 0.019$ & $0.377 \pm 0.017$ & 1.277 & 2.689 \\ 
& 3-day & Double Log-normal & $0.551 \pm 0.188$ & $0.039 \pm 0.008$ & $0.352 \pm 0.006$ & $0.091 \pm 0.012$ & $0.393 \pm 0.033$ & 1.707 & 5.219 \\ 
4FGL J1443.9+2501 & 1-month & Log-normal & - & $0.058 \pm 0.005$ & $0.355 \pm 0.007$ & - & - & 1.075 & 0.743 \\ 
& 7-day & Double Log-normal & $0.853 \pm 0.214$ & $0.080 \pm 0.015$ & $0.370 \pm 0.032$ & $0.025 \pm 0.021$ & $0.274 \pm 0.013$ & 1.001 & 0.906 \\ 
& 3-day & Double Log-normal & $0.397 \pm 0.107$ & $0.042 \pm 0.009$ & $0.315 \pm 0.008$ & $0.125 \pm 0.014$ & $0.383 \pm 0.020$ & 0.957 & 3.701 \\ 
4FGL J1512.8-0906 & 1-month & Double Log-normal & $0.332 \pm 0.073$ & $0.044 \pm 0.005$ & $0.352 \pm 0.006$ & $0.014 \pm 0.002$ & $0.386 \pm 0.002$ & 0.833 & 7.333 \\ 
& 7-day & Double Log-normal & $0.734 \pm 0.129$ & $0.026 \pm 0.003$ & $0.386 \pm 0.002$ & $0.086 \pm 0.024$ & $0.374 \pm 0.014$ & 1.088 & 6.789 \\ 
& 3-day & Double Log-normal & $0.345 \pm 0.067$ & $0.092 \pm 0.012$ & $0.392 \pm 0.005$ & $0.034 \pm 0.002$ & $0.387 \pm 0.002$ & 1.537 & 7.508 \\ \hline
\end{tabular}
}
\end{table*}


\subsubsection{Correlation Study and Two-Flux-State Hypothesis}

In our study, we examined the relationship between $\gamma$-ray flux and spectral index to identify characteristic trends in blazar variability. Most sources in our sample exhibit a ``harder-when-brighter" trend, where the spectral index decreases (hardens) with increasing flux. This behavior, observed across all time binnings (3-day, 7-day, and 30-day) in nine sources — 4FGL J0348.5-2749, 4FGL J0739.2+0137, 4FGL J0904.9-5734, 4FGL J1159.5+2914, 4FGL J1224.9+2122, 4FGL J1256.1-0547, 4FGL J1310.5+3221, 4FGL J1422.3+3223, and 4FGL J1443.9+2501 — highlights the role of efficient particle acceleration processes like shocks, turbulence, or magnetic reconnection within the relativistic jet. Here the particles get accelerated faster than they cool resulting in the harder spectral indices during periods of rising flux. This suggests that in high-flux states, the contribution of accelerated particles dominates, while in low-flux states, cooling processes become more significant \citep{1998A&A...333..452K}.

Conversely, two sources — 4FGL J0221.1+3556 and 4FGL J1512.8-0906 — display a ``softer-when-brighter" trend, where the spectral index increases (softens) as flux rises. This behavior is likely linked to enhanced radiative cooling at higher flux levels, particularly in environments with strong external photon fields such as the broad-line region or dusty torus. In these environments, accelerated particles lose energy more rapidly through processes like inverse Compton scattering, shifting the emission to lower energies and softening the spectrum.

In our analysis, we found that the Spearman rank correlation coefficients for most of the sources became stronger as we moved to larger time bins. This indicates that the overall correlation—whether positive or negative—between flux and photon index becomes more pronounced in most sources when observed over longer timescales. Table \ref{table:correlation} summarizes the correlation coefficients calculated for different time bin sizes. This result suggests that, on shorter timescales, flux variations are influenced by both index and normalization (norm) fluctuations. As the time bin size increases, the averaging of flux over longer durations reduces the effect of norm fluctuations, effectively smoothing out their variations. This effect enhances the strength of the correlation between flux and photon index, making it more prominent on longer timescales \citep{2017MNRAS.470.3283S}.

Several studies, including those by \citet{2016ApJ...822L..13K} and \citet{2020MNRAS.491.1934K} have proposed a ``two-flux-state hypothesis” for blazars, suggesting that blazars may exhibit distinct low and high flux states. Our results lend support to this hypothesis, as double lognormal profiles were observed consistently in both flux and index distributions within the 3-day binning and mostly in 7-day binning, indicating the presence of multiple emission states. These double-distribution profiles, which tend to become insignificant in the 30-day bin, likely reflect that shorter timescales are more effective at capturing distinct states, while longer timescales smooth out these variations.

To further investigate the two-flux-state hypothesis, we analyzed the intersection of centroids from the double lognormal profiles fitted to the flux and index distributions. For sources exhibiting the ``harder when brighter" trend, the higher flux component corresponds to a lower spectral index, as indicated by the same color in the plots. Conversely, for sources following the ``softer when brighter" trend, the higher flux component is associated with a higher spectral index, also represented by the same color. By examining the intersection points of these components (marked by cross symbols in the plots), which lie above and below the best-fit line for most sources, we observe a clear pattern. The points above the line (shown in grey) and those below the line (shown in light cyan) seem to represent two distinct flux states. This observation provides additional evidence supporting the hypothesis of two separate flux states (See Figure \ref{fig:fi3day} for 3-day, Figure \ref{fig:fi7day} for 7-day and Figure \ref{fig:fi30day} for 30-day).

Additionally, we calculated the mean values for data points lying above and below the best-fit line and included these in the plots, represented by orange and black filled circles, respectively. In most cases, the mean of the lower points aligns with the lower flux component at the intersection point within 1$\sigma$ error band. However, the mean of the upper points tends to deviate from the higher flux component at the intersection point above the best-fit line. This indicates that the lower flux component is primarily influenced by index variations, while the higher flux component likely involves fluctuations from both the index and norm factors. These results suggest that flux variations are not solely driven by index changes, which depend on acceleration and cooling timescales, but are also affected by norm factors, such as the magnetic field and Doppler factor.

We also plotted the corresponding flux and index values at times when the source was detected in the VHE. In all cases, we found that these points fall within the higher flux and lower index range, which is consistent with findings suggesting that VHE detections are more probable during high-flux, hard-spectrum states. This aligns with theoretical predictions and prior empirical observations, highlighting that VHE detection is often associated with the most energetic states of the blazar, providing insights into the nature of high-energy particle acceleration and emission mechanisms at work in these extreme conditions.


\begin{table*}
\caption{Spearman rank correlation coefficient ($r_s$) and p-value ($P_s$) for various sources with 3-day, 7-day, and 30-day binning intervals.}
\label{table:correlation}
\centering
\begin{tabular}{ccccccc}
\hline
\textbf{Source} & \multicolumn{2}{c}{\textbf{3-day}} & \multicolumn{2}{c}{\textbf{7-day}} & \multicolumn{2}{c}{\textbf{30-day}} \\
 & \( r_s \) & \( P_s \) & \( r_s \) & \( P_s \) & \( r_s \) & \( P_s \) \\
\hline \hline
4FGL J0221.1+3556 & 0.28 & 6.25e-10 & 0.24 & 5.77e-07 & 0.34 & 2.49e-06 \\
4FGL J0348.5-2749 & -0.40 & 2.73e-26 & -0.47 & 3.10e-21 & -0.62 & 7.54e-14 \\
4FGL J0739.2+0137 & -0.19 & 3.74e-03 & -0.20 & 2.70e-03 & -0.32 & 8.73e-05 \\
4FGL J0904.9-5734 & -0.14 & 1.24e-04 & -0.15 & 2.19e-03 & -0.21 & 7.23e-03 \\
4FGL J1159.5+2914 & -0.31 & 2.42e-25 & -0.40 & 6.02e-25 & -0.59 & 8.01e-18 \\
4FGL J1224.9+2122 & -0.18 & 6.44e-07 & -0.30 & 1.97e-10 & -0.29 & 4.81e-04 \\
4FGL J1256.1-0547 & -0.11 & 5.19e-06 & -0.19 & 1.83e-07 & -0.26 & 2.79e-04 \\
4FGL J1310.5+3221 & -0.19 & 1.18e-05 & -0.29 & 5.82e-06 & -0.40 & 1.01e-06 \\
4FGL J1422.3+3223 & -0.27 & 1.14e-09 & -0.34 & 3.06e-09 & -0.44 & 1.62e-05 \\
4FGL J1443.9+2501 & -0.10 & 1.18e-01 & -0.27 & 8.01e-04 & -0.05 & 6.35e-01 \\
4FGL J1512.8-0906 & 0.02 & 4.02e-01 & 0.10 & 4.15e-03 & 0.29 & 3.96e-05 \\
\hline
\end{tabular}
\end{table*}


\subsubsection{Comparison of VHE and Non-VHE FSRQs}

The \emph{Fermi} LCR includes sources from the Fermi 4FGL-DR2 catalog, specifically selecting those with variability indices greater than 21.67. This variability index, which reflects the fractional variability (\(\delta F / F\)) over yearly timescales, resulted in a sample of 1525 variable sources. Among these, approximately $38\%$ are classified as FSRQs, making them the largest category of variable sources detected in the catalog.

Following the previously described data cuts and incorporating the results of the AD test, we fitted each FSRQ with the best distribution based on the reduced \(\chi^2\) value. Based on the results from the previous section, where all 11 VHE FSRQs were shown to exhibit double lognormal distributions in both flux and index under finer time binning (3-day), we extended our analysis to identify additional FSRQs with similar distribution characteristics. Out of the entire sample of non-VHE FSRQs, we found that 99 sources also confirmed to a double lognormal distribution for both flux and index in the 3-day binning. This group of 99 non-VHE FSRQs provides a basis for comparing with the VHE FSRQs. By analyzing their flux and index distribution parameters, we can explore possible differences in their emission properties, which may point to distinct underlying physical processes in VHE versus non-VHE FSRQs. Figure \ref{fig:comp} provides a comparison of the statistical parameters from the double lognormal fits for VHE and Non-VHE FSRQs.

The top two panels illustrate the relationship between the mean flux ($\mu_1$-Fux and $\mu_2$-Flux) and the mean photon index ($\mu_1$-Index and $\mu_2$-Index) for VHE and Non-VHE FSRQs, compared to the overall average behavior of the FSRQ population, as indicated by the best-fit lines. VHE FSRQs exhibit slightly harder spectra compared to the average indices of FSRQs at similar flux levels. This indicates that, while VHE FSRQs do not occupy distinct regions, they show subtle differences in spectral behavior, consistent with harder spectra.

The bottom two panels compare the flux dispersions ($\sigma_1$-Flux and $\sigma_2$-Flux) with the index dispersions ($\sigma_1$-Index and $\sigma_2$-Index). In the first component ($\sigma_1$-Flux vs. $\sigma_1$-Index), VHE FSRQs show lower index dispersions, indicating reduced variability and more stable spectra compared to the average FSRQs. In contrast, the second component ($\sigma_2$-Flux vs. $\sigma_2$-Index) shows higher index dispersions for VHE FSRQs, reflecting greater variability in this component. These differences suggest that VHE FSRQs exhibit more stability in one component while showing higher variability in the other, possibly due to differences in emission processes.

\begin{figure*}
\centering
\includegraphics[scale=0.45,angle=0]{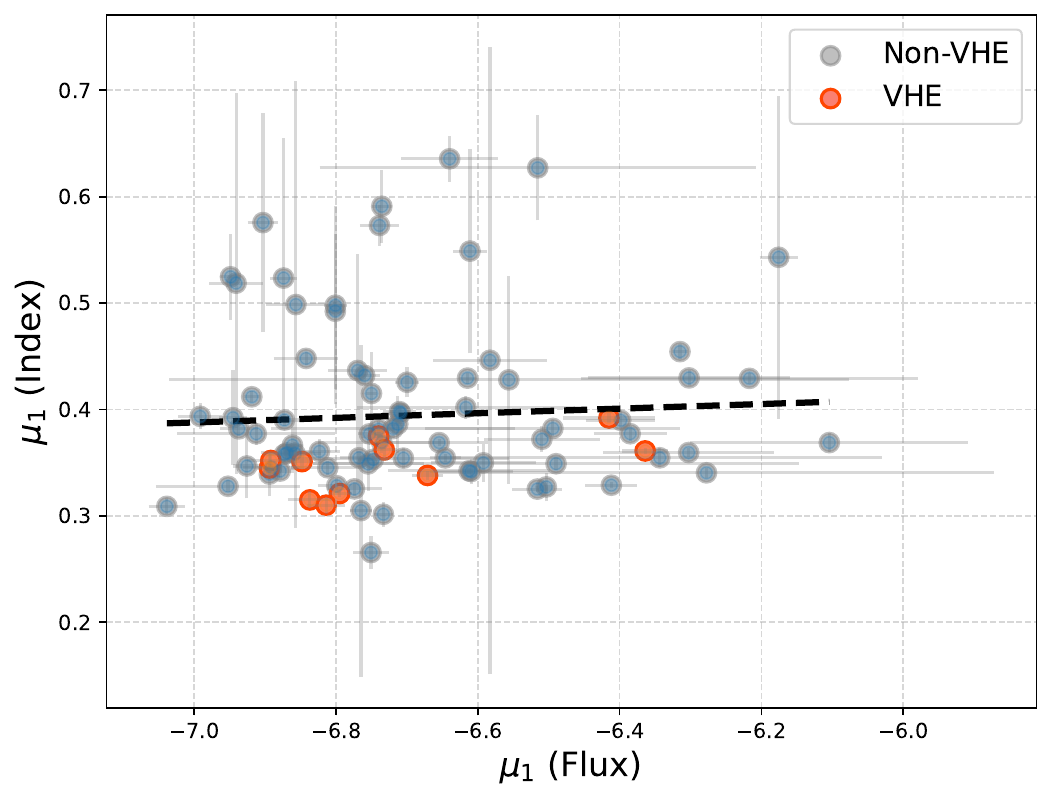}
\includegraphics[scale=0.45,angle=0]{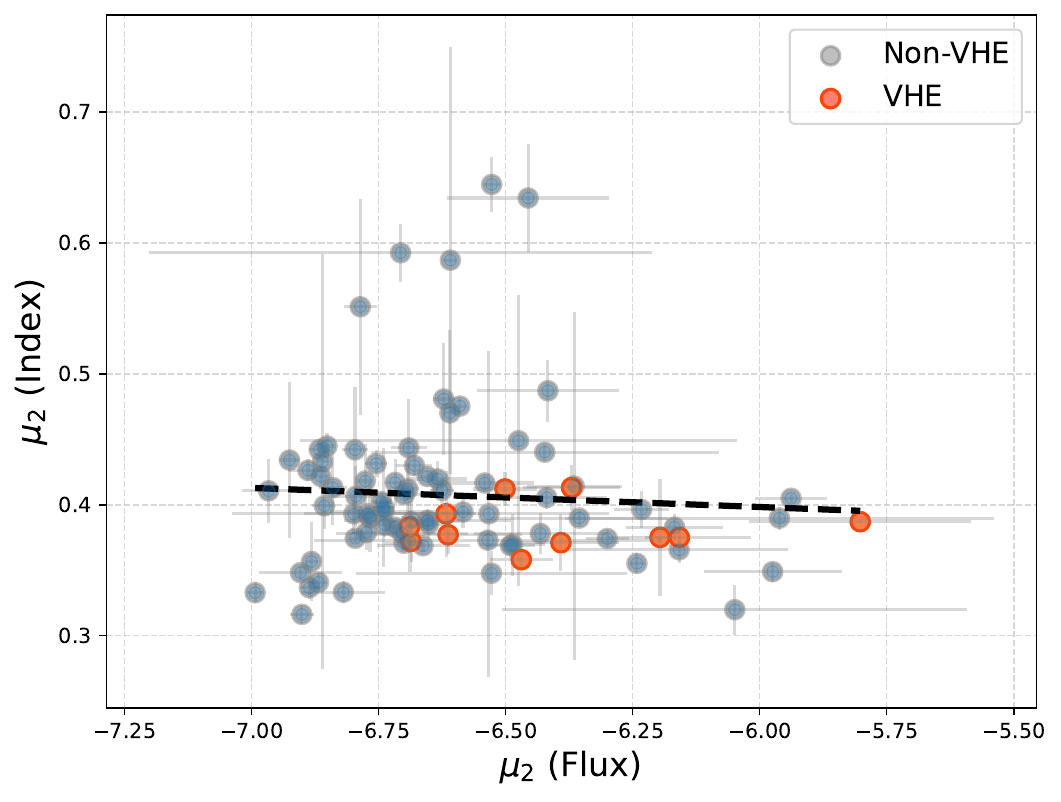}
\includegraphics[scale=0.45,angle=0]{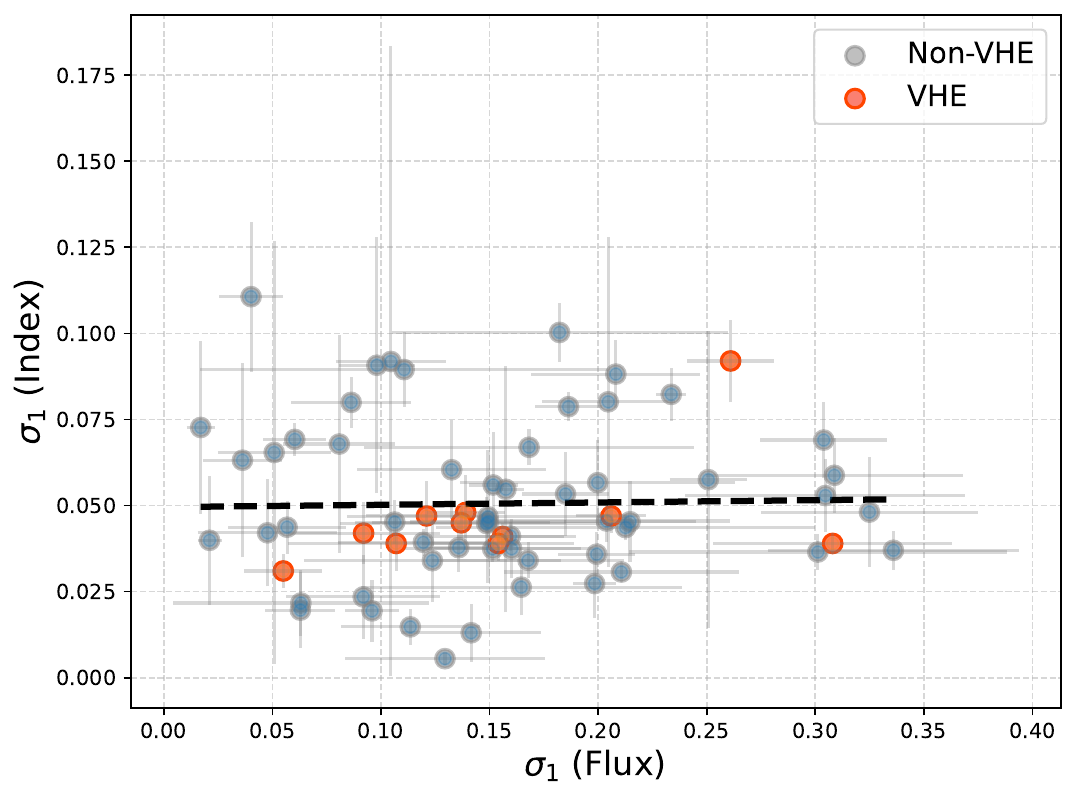}
\includegraphics[scale=0.45,angle=0]{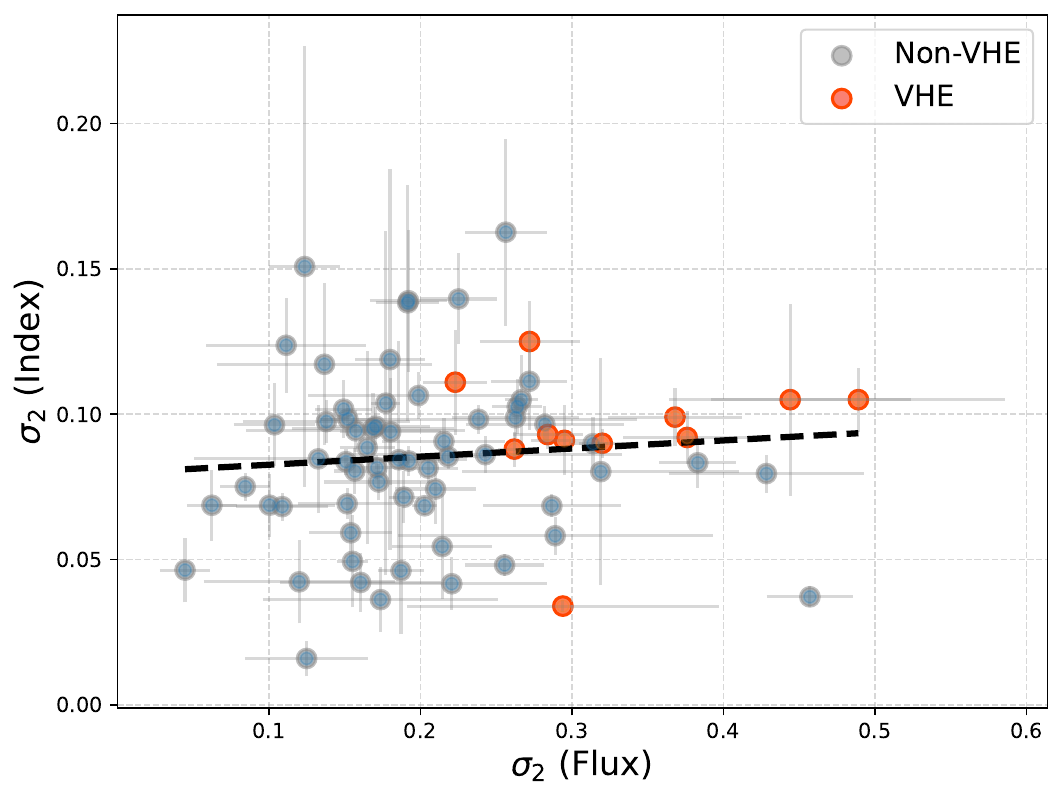}\\
\caption{Comparison of parameters between VHE and Non-VHE FSRQs. The panels depict the relationships between flux and index distribution parameters. The black dotted line in each plot represents the best-fit line for all data points.}
\label{fig:comp}
\end{figure*}

\section{Summary}
\label{sec:summary}

This study investigates the $\gamma$-ray variability and statistical properties of 11 VHE FSRQs using light curves derived from \emph{Fermi}-LAT observations in the 0.1–100 GeV energy range. By using different time bins of 3 days, 7 days, and 30 days, the analysis captures variability trends across both short and long-term timescales, offering critical insights into the highly dynamic behavior of these sources. The light curves reveal significant $\gamma$-ray flux variability and spectral changes in all 11 VHE FSRQs. During periods of VHE activity, the \emph{Fermi} $\gamma$-ray flux increased in most cases, suggesting a connection between enhanced particle acceleration and the emergence of VHE emissions.

The statistical distributions of $\gamma$-ray flux and photon index revealed that double log-normal profiles were the most common in finer binnings, particularly in the 3-day and 7-day intervals. This aligns with the ``two-flux-state hypothesis", which suggests that blazars can exhibit distinct low and high flux states. Further analysis of the centroids of these double log-normal profiles provided additional support for this hypothesis. As the time bin size increased, the distributions transitioned from double log-normal to simpler forms, reflecting the smoothing effects of longer-term variability over short-term fluctuations.

Correlation analysis between $\gamma$-ray flux and photon index revealed diverse behaviors among the sources. Nine of the 11 VHE FSRQs showed a ``harder-when-brighter" trend, where the spectral index hardened as the flux increases, indicating more efficient particle acceleration over cooling during high-energy states. In contrast, two sources displayed a ``softer-when-brighter" trend, likely due to strong radiative cooling effects from interactions with dense external photon fields. Furthermore, the correlation strength between flux and index increased with larger time bin sizes in most cases, suggesting that shorter timescales are sensitive to variations in both the index and normalization factors, while at longer timescales normalization variations get minimized. Additionally, When compared to non-VHE FSRQs, the VHE sources are found to have harder spectra and show some differences in their variability patterns. 

In conclusion, this study provides valuable insights into the $\gamma$-ray variability and statistical properties of VHE FSRQs. The results suggest that the high-energy emissions of these sources are driven by complex processes, including particle acceleration, cooling effects, and normalization factors. Future work involving a larger sample of sources and multi-wavelength observations could offer deeper insights into the physical mechanisms responsible for the variability and high-energy emissions in these extreme blazars, aiding in the identification of additional VHE candidates and the refinement of theoretical models of $\gamma$-ray emission.

\section*{Acknowledgements}

The authors thank the anonymous referee for valuable comments and suggestions. We have used the data from Fermi Light Curve Repository (LCR), which is currently maintained by Daniel Kocevski (NASA MSFC) with support from Janeth Valverde, Simone Garrappa, Michela Negro, Jean Ballet, and Benoit Lott. The website interface was designed by Daniel Kocevski, incorporating tools such as Olaf Frohn's D3-Celestial sky map code, the Highcharts JavaScript plotting library, and the Bootstrap toolkit. The development of the LCR has been funded in part through the Fermi Guest Investigator Program. ZM acknowledges the financial support provided by the Science and Engineering Research Board (SERB), Government of India, under the National Postdoctoral Fellowship (NPDF), Fellowship reference no. PDF/2023/002995. SA is thankful to the MOMA for the MANF fellowship (No.F.82-27/2019(SA-III)). ZS is supported by the Department of Science and Technology, Govt. of India, under the INSPIRE Faculty grant (DST/INSPIRE/04/2020/002319). ZM express  gratitude to the Inter-University Centre for Astronomy and Astrophysics (IUCAA) in Pune, India, for the support and facilities provided.

\section*{Data Availability}

This research has used $\gamma$-ray observations from Fermi Light Curve Repository (LCR) which can be accessed at \url{https://fermi.gsfc.nasa.gov/ssc/data/access/lat/LightCurveRepository/}.



\bibliographystyle{mnras}
\bibliography{example} 





\bsp	
\label{lastpage}
\end{document}